\documentclass[12pt]{article}
\usepackage{geometry}
\AtBeginDocument{%
  }

\usepackage{graphicx}
\usepackage[colorlinks=true, linkcolor=black, citecolor=black, urlcolor=black]{hyperref}
\hypersetup{breaklinks=true}

\begin{document}


\title{A Survey of AI Music Generation Tools and Models}
\author{%
    Yueyue Zhu\thanks{Department of Computer Science, Metropolitan College, Boston University, Boston, MA, USA. Email: \texttt{yueyuez@bu.edu}} \and
    Jared Baca\thanks{Department of Contemporary Writing and Production, Berklee College of Music, Boston, MA, USA. Email: \texttt{jbaca@berklee.edu}} \and
    Banafsheh Rekabdar\thanks{Department of Computer Science, Maseeh College of Engineering and Computer Science, Portland State University, Portland, OR, USA. Email: \texttt{rekabdar@pdx.edu}} \and
    Reza Rawassizadeh\thanks{Department of Computer Science, Metropolitan College, Boston University, Boston, MA, USA. Email: \texttt{rezar@bu.edu}}
}
\date{\today} 

\maketitle

\begin{abstract}
  In this work, we provide a comprehensive survey of AI music generation tools, including both research projects and commercialized applications. To conduct our analysis, we classified music generation approaches into three categories: parameter-based, text-based, and visual-based classes.
  Our survey highlights the diverse possibilities and functional features of these tools, which cater to a wide range of users, from regular listeners to professional musicians. We observed that each tool has its own set of advantages and limitations. As a result, we have compiled a comprehensive list of these factors that should be considered during the tool selection process.
  Moreover, our survey offers critical insights into the underlying mechanisms and challenges of AI music generation.
\end{abstract}

\section{Introduction}
Music is an integral part of human culture that has evolved significantly over the centuries, adapting to different cultures, styles, and technologies \cite{denisova2017evolution}. Music generation by models has also undergone a paradigm shift with the advancements in AI and machine learning \cite{hernandezolivan2021music}. Artificial intelligence (AI) music generation tools provide musicians and composers with new and innovative ways to create music, which has not only facilitated the expression of users' musical intent but also had a significant impact on their creative ownership and confidence in their ability to collaborate with AI technology \cite{Carnovalini_2020, Louie_2020,zhao2022review}. These tools \cite{google_ai, magenta_github, huang2022mulan, MusicLM_paper, Di_2021} use machine learning algorithms to learn from large music datasets and aim to generate new music compositions that are indistinguishable from human-generated music. 

The advent of deep neural networks, a.k.a, deep learning, since 2012, has revolutionized several computer science disciplines \cite{lecun2015deep}, including AI music generation.   
Several deep learning models can generate short-term note sequences, but creating longer melodies has been made possible through the development of recent neural network architectures, such as MusicVAE \cite{roberts2019hierarchical} and TransformerVAE \cite{9054554}, and generative models such as Denoising Diffusion Probabilistic Models \cite{hernandezolivan2021music}. However, these models generate longer polyphonic melodies that do not necessarily follow a central motif and may need a sense of direction. Deep learning models have been used for harmonization, generating the harmony that accompanies a given melody, and style transfer techniques have been used to transform music in a determined style into another style. Briot, Jean-Pierre et al.\cite{Briot2020} discussed the limitations in directly applying deep learning to music generation, including the lack of creativity and control and the need for human interaction in the composition process. Data-driven AI models sometimes have the tendency to produce variations of existing patterns instead of completely original compositions\cite{Briot2020}. This constraint stems from their reliance on training data, which inherently bounds their creative output. 

In this survey, first, we explain the fundamental terms common in music generation, which are also applied in AI-generated music. Then, we will explore the current state of AI music generation tools and models, evaluating their functionality and discussing their limitations. Finally, by analyzing the latest tools and techniques, we aim to provide a comprehensive understanding of the potential of AI-based music composition and the challenges that must be addressed to improve their performance.

This survey aims to provide an overview of AI music generation tools and models, their capabilities, and their limitations. We start by explaining concepts to readers who are not familiar with music composition. Then, we describe our methods. In particular, we start by explaining our data collection approach. Then, we list traditional approaches without neural networks and generating music. Next, we will examine the common AI-based music generation tools available today. These tools are open-source and have been used by several researchers and developers to create AI-generated music. However, only some of the models we reviewed were open-source, and in such cases, we relied on official demos or explanations for comparison.
The scope of this work is limited to AI music generation tools and models that employ machine learning algorithms to create music. We will not cover the broader applications of AI in music, such as music classification, recommendation systems, and music analysis.

\section{Music Composition Concepts}
In this section, we will explore fundamental concepts that contribute to the structure and organization of a musical composition. Understanding their interplay is crucial for developing AI-generated music tools.

\begin{figure}[h]
    \centering
    \includegraphics[width=\textwidth]{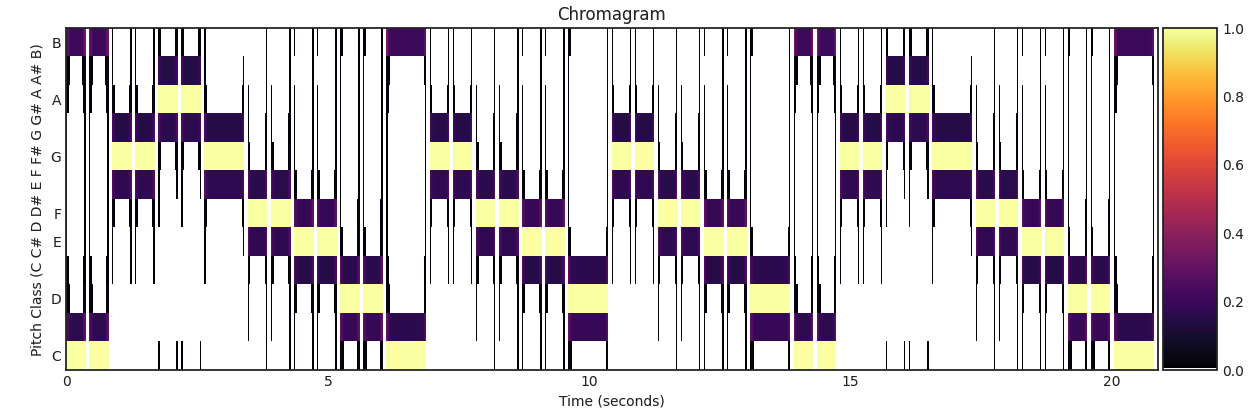}
    \caption{Chromagram of "Twinkle Twinkle Little Star", illustrating pitch class energy across time.}
    \label{fig:chromagram}
\end{figure}

\textbf{Tone} represents a sound with a definite pitch, characterized by its frequency, amplitude, and timbre \cite{rossing2002science}. It serves as the fundamental unit in music composition, allowing the creation of melodies, chords, and other musical structures.\\
\textbf{Pitch (Key)} signifies the perceived frequency of a sound, determining its position on the high-low spectrum \cite{roederer2013physics}. A musical piece is typically composed in a specific pitch, establishing the tonal center and governing the relationships between the notes \cite{laitz2012complete}. \\
\textbf{Timbre} often referred to as the tonal color or quality of a sound, is the characteristic that distinguishes different sound sources even when they have the same pitch and volume\cite{RISSET1999113, sethares2005tuning}.\\ 
\textbf{Harmony} refers to the simultaneous combination of different pitches or tones that generate a pleasing auditory sensation for the listener \cite{piston1987harmony}.\\
\textbf{Chords} refers to groups of notes that played concurrently, form the foundation of harmony in music.\\
\textbf{Tempo} indicates the speed at which a musical composition is performed, typically measured in beats per minute (BPM) \cite{london2004hearing}.
Tempo can significantly influence a piece's mood and emotional impact, with faster tempos often associated with excitement or energy and slower tempos linked to calmness or sadness \cite{sloboda2005exploring}.  AI-generated music tools can strategically adjust tempo to evoke specific emotions in listeners, tailoring the generated compositions to fit desired moods or feelings. \\
\textbf{Volume} denotes the perceived loudness of a sound, which is closely related to its amplitude or intensity. It is a scalar quantity that represents the magnitude of the acoustic energy being transmitted, typically measured in decibels (dB)\cite{moser2009engineering}.\\
\textbf{Style} encompasses the distinctive characteristics and techniques utilized by a composer or performer, thereby shaping the unique identity of their musical creations \cite{meyer1989style}. When style is applied to AI-generated music tools, analyzing and learning from existing musician-created music enables the emulation of styles from different composers or genres, consequently generating new compositions that reflect the unique artistic traits of the artist or historical era.\\
\textbf{Chorus} refers to a recurrent section within a song, often featuring a memorable melody and lyrics that convey piece's central theme \cite{covach2005what}.\\
\textbf{Polyphonic music}  refers to music that consists of multiple independent melody lines played or sung simultaneously. These melody lines interact with each other to create harmonies, counterpoints, and textures that are richer and more complex than monophonic music, which consists of a single melody line.\\
\textbf{MIDI} (Musical Instrument Digital Interface) is a standard protocol for communication between electronic musical instruments, computers, and other digital devices. MIDI enables the exchange of musical information, such as notes, velocities, and control messages, between different devices and software applications. It allows musicians and producers to control and synchronize different instruments and devices, and to record and edit musical performances with precision and flexibility. \\
\textbf{Key Velocity} also known as key strike velocity or keystroke velocity, is a measurement of how forcefully a key is pressed on a MIDI keyboard or other MIDI controller. This value is usually expressed as a number between 0 and 127, with 0 indicating that the key was not pressed at all, and 127 indicating that the key was pressed with maximum force.\\
\textbf{ABC or abc notation}\cite{abcnotation} is a shorthand notation system for writing music using ASCII characters. Developed in the 1970s and 1980s, it is commonly used in the Celtic and folk music traditions to share and distribute traditional music. It uses letters and symbols to represent musical notes, rhythms, and other elements, making it an easy-to-learn and widely used method for sharing music in digital formats such as  over the Internet.\\
\textbf{Pianoroll} is an interface in Digital Audio Workstations (DAWs) \cite{leider2004digital} enabling manipulation of Musical Instrument Digital Interface (MIDI) data. It utilizes a grid where the X-axis denotes time and the Y-axis represents pitch. The duration and intensity (velocity) of notes are adjustable, making them integral to the structure of musical compositions and vital for developing AI-generated music tools.\\
\textbf{Chromagram} is a visualization that illustrates the intensity of various pitches, in a musical piece, over time. As demonstrated in Figure \ref{fig:chromagram}, each pitch class (C to B) corresponds to a unique row on the y-axis and the x-axis represents time. The color at each point in this 2D space reveals the energy level of a pitch class at the given time,  brighter colors indicating higher energies. Figure \ref{fig:chromagram} Chromagram depicts the melody of "Twinkle Twinkle Little Star".\\
\textbf{Accompaniment} in music refers to the supportive, often harmonic, elements that provide a backdrop for the main melody or theme of a song. \\

\subsection{Interplay of concepts in AI-generated Music}

AI-generated music tools can produce compositions with coherent and aesthetically pleasing progressions by incorporating knowledge of harmony and chords.  Additionally, identifying patterns in choruses of popular songs allows AI-generated music tools to create new songs with catchy and memorable melodic hooks that resonate with listeners.

In conclusion, understanding the interplay of these fundamental musical concepts is vital for developing sophisticated AI-generated music tools capable of producing human-like compositions which are emotionally expressive and contextually appropriate.  By effectively modeling and utilizing these concepts, an AI model can create compositions that potentially enrich the musical landscape with innovative structures and motifs, demonstrating a promising avenue for the future of music generation and technology. Furthermore, this knowledge bridges the gap between traditional music creation and AI-generated music, facilitating interdisciplinary collaboration and driving advancements in music technology.

\section{Methods}

\subsection{Data Collection}
To create a comprehensive list of AI music generation tools and models, we employed a keyword search method across several platforms, along with the assistance of ChatGPT v3.5, ChatGPT v4, and Bard \footnote{https://bard.google.com} to refine the keyword search list and web resources. To identify the list of keywords, first, three authors of the paper compile a list of keywords separately, and then they accumulate their results.  Next, we use the listed Large Language Models (LLMs) to refine the list of keywords and refine the list of web resources to search for these keywords.

The list of web resources: \emph{Google Search, Google News, Bing News, Google Scholar, Twitter, GitHub, YouTube, Reddit \footnote{https://reddit.com}, Hacker News \footnote{https://news.ycombinator.com}}, and \emph{Huggingface} \footnote{https://huggingface.co}

The list of keywords: \emph{AI music, AI music generation, Diffusion music generation, Neural Networks Music Generation,  Machine Learning Music, Music Generation Models, Music Generation Algorithms, Music AI, Music Technology, Computer-generated Music,  Deep Learning Music.}

The prompt we have used on our LLM platform is as follows:
\textit{I am searching for music generated AI in the following platforms; [web resources] by using the following keywords; [keywords]]. Do you have any platform or keyword recommendation that I missed?}

\subsection{Taxonomy of Music Generation Tools}\label{sec:categorization}

\begin{figure}[h] 
    \centering
    \includegraphics[width=1.0\textwidth]{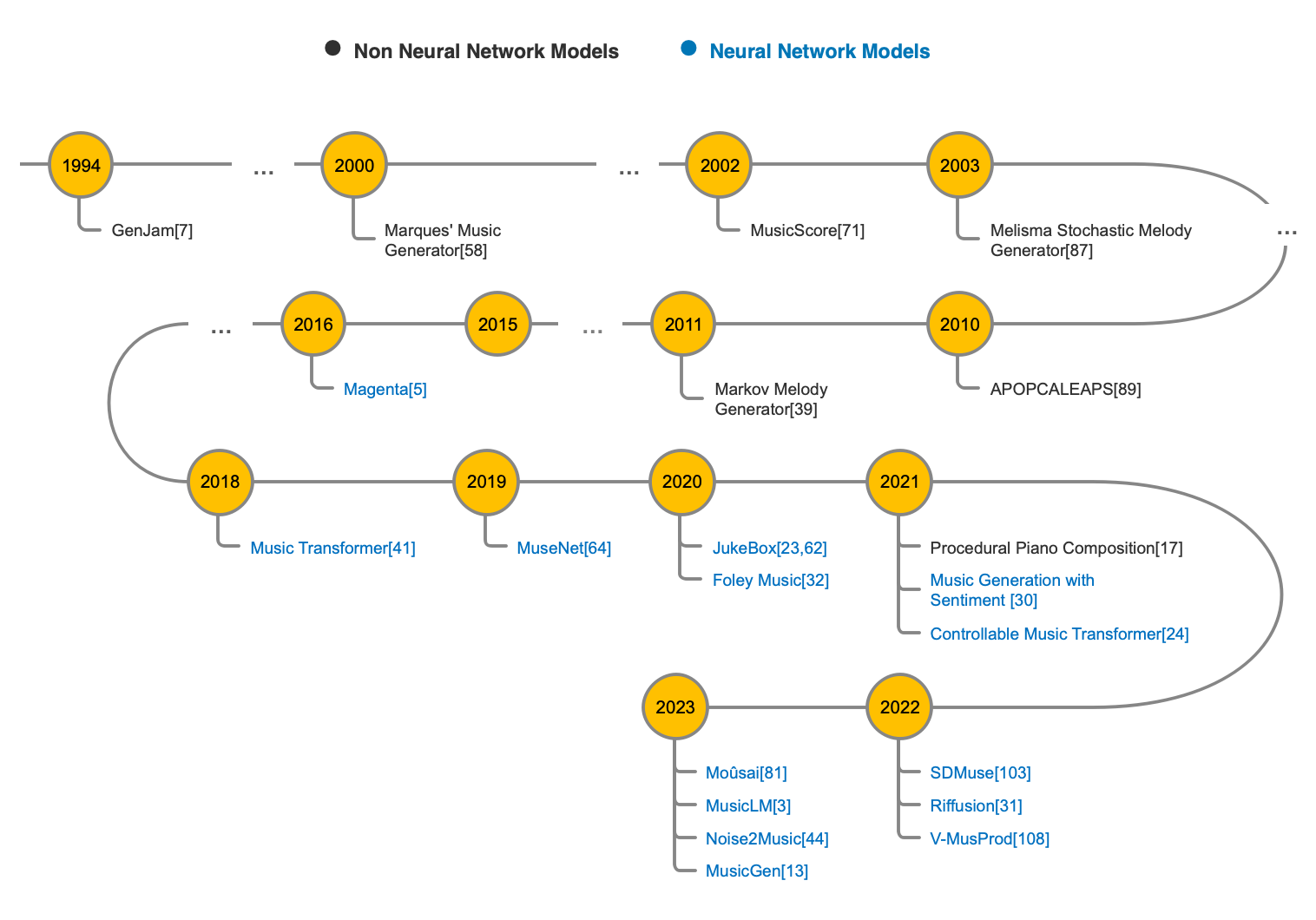} 
    \caption{Timeline of Music Generation Models}
    \label{fig:timeline} 
\end{figure}

We provide a visual representation of the progression of these music generation models from their inception to the present day in Figure \ref{fig:timeline}. The timeline traces the development from the early non-neural-network methods to the latest AI-enabled, parameter-free models, clearly illustrating the significant advancements in music generation technology over the years.

Traditionally, music generation AI tools use a variety of methods, including Markov chain\cite{Gautam2018MusicCW}, rule-based models \cite{rule-based-generation, APOPCALEAPS}, and evolutionary algorithms\cite{zhao2022review, Farzaneh2019, evolutionary_2011}. These methods are typically parameter-based, necessitating human input of relevant parameters or configurations to guide the music-generating process.

Later models have relied on neural networks. Since the quality of generated music via the neural network is significantly better than traditional methods, we separate their explanation in this survey. We classify neural network models into two categories: parameter-based and parameter-free models. Parameter-based models are those that require specific input parameters, such as the desired 'tempo' or 'key', to generate music. On the other hand, parameter-free models do not require any specific input parameters, and they are further subdivided into two categories: prompt-based and visual-based models. Prompt-based models allow users to input descriptive texts as prompts to generate music, while visual-based models use visual inputs such as images or videos.

\begin{figure}[h] 
\centering
\includegraphics[width=0.9\textwidth]{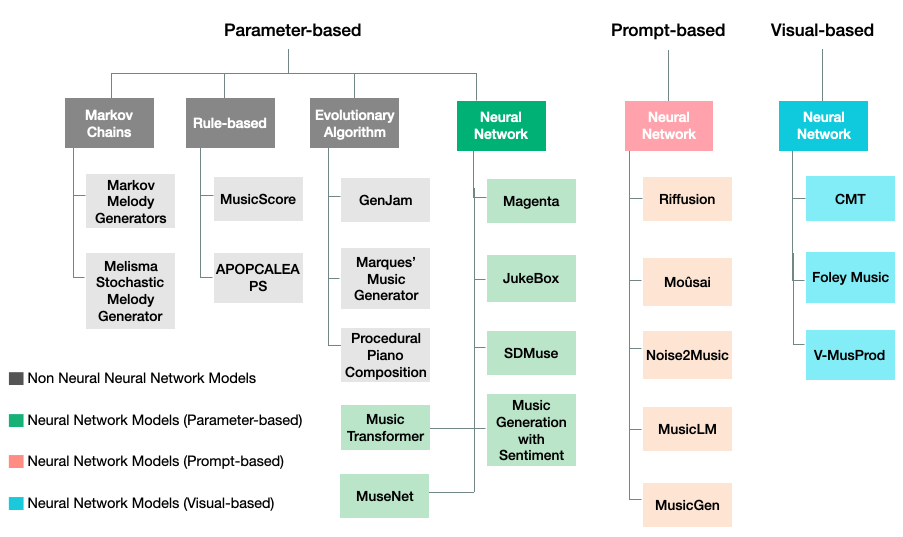} 
\caption{A Hierarchical Taxonomy of Music Generation Models}
\label{fig:tree} 
\end{figure}

Furthermore, Figure \ref{fig:tree} presents a hierarchical taxonomy of the music generation models, which further elucidates the categorizations and subcategories of these models. This tree structure allows a more granular understanding of the differences and relationships between the various models, as well as their evolution over time.

\begin{table}[htbp]
\centering
\caption{Comparison of non-neural network music generation tools}
\small
\begin{tabular}{|p{2.8cm}|p{5.5cm}|p{2.4cm}|p{3cm}|}
\hline
\textbf{Tool/Method} & \textbf{Main Function} & \textbf{Non-Neural Network Model} & \textbf{Classification} \\
\hline
Markov Melody Generator \cite{SimoneHill2023} & Analyzes patterns in existing music to create new compositions & Markov Chains\cite{Shapiro2021,MarkovProcess,Ramanto2017MarkovCB} & Parameter-based\\
\hline
Melisma Stochastic Melody Generator \cite{Melisma2003} & Generates melody sequences using stochastic processes & Markov Chains\cite{Melisma2003} & Parameter-based\\
\hline
MuseScore \cite{MuseScore2011} & Allows composition based on user-defined patterns & Rule-Based Systems\cite{MuseScore2011} & Parameter-based\\
\hline
APOPCALEAPS \cite{APOPCALEAPS} & Generates pop music based on chance rules & Rule-Based Systems\cite{rule-based-generation,APOPCALEAPS} & Parameter-based \\
\hline
GenJam \cite{Biles1994} & Uses selection, mutation, and crossover of musical sequences to generate new music & Genetic Algorithm\cite{Biles1994} & Parameter-based \\
\hline
Procedural Piano Composition \cite{shu28011} & Evolves random music pieces using mood templates until their mood aligns with the template & Genetic Algorithm\cite{shu28011} & Parameter-based \\
\hline
Marques' Music Generator \cite{marques2000music} & Uses melodic and musical theory concepts as a fitness function in a genetic algorithm & Genetic Algorithm\cite{marques2000music} & Parameter-based \\
\hline
\end{tabular}
\label{table1}
\end{table}

\begin{table}[htbp]
\centering
\caption{Comparison of neural network music generation tools}
\small
\begin{tabular}{|p{2.8cm}|p{3.5cm}|p{2.4cm}|p{2.7cm}|p{2.3cm}|}
\hline
\textbf{Tool Name} & \textbf{Main Function} & \textbf{Access Interface} & \textbf{Neural Network Model} & \textbf{Classification} \\
\hline
Magenta\cite{google_ai} & Melody, drum, piano, chords generation using deep learning & Plug-in, command-line, Python, jscript & RNN-LSTM\cite{LSTM_intro, Sherstinsky_2020}), VAE\cite{kingma2022autoencoding}, GANSynth\cite{engel2018gansynth}, WaveNet\cite{oord2016wavenet}, Transformer networks\cite{vaswani2017attention} & Parameter-based \\
\hline
Jukebox\cite{jukebox, openai_2020} & Music generation with vocal synthesis & Python & VQ-VAE\cite{oord2018neural} & Parameter-based \\
\hline

Music Generation with Sentiment\cite{ferreira2021learning} & Music generation with sentiment analysis & Python & mLSTM\cite{krause2017} & Parameter-based \\
\hline
MuseNet\cite{openai-musenet} & Music generation with style transfer capabilities & Web-based & Transformer\cite{vaswani2017attention} (Decoder-only) & Parameter-based \\
\hline
Music Transformer\cite{huang2018music} & Music generation with long-term structure & Python & Transformer\cite{vaswani2017attention} & Parameter-based \\
\hline
SDMuse\cite{sdmuse} & Music editing and generation framework & Not released & SDE\cite{song2021scorebased}, Gaussian Diffusion Model\footnote{https://github.com/openai/guided-diffusion} & Parameter-based \\

\hline
Riffusion\cite{Forsgren_Martiros_2022} & Audio clip generation from text and images & Web User Interface & Stable Diffusion v1.5\cite{stable_diffusion} & Prompt-based \\
\hline
Moûsai\cite{schneider2023mousai} & Generative music from textual prompts & Python & Diffusion Magnitude Autoencoder (DMAE)\cite{DMAE} & Prompt-based \\
\hline
Noise2Music\cite{huang2023noise2music} & Generate high-quality music audio from text prompts & Not released & Diffusion Model\cite{ratcliff2004diffusion, pmlr-v37-sohl-dickstein15} & Prompt-based \\
\hline
MusicLM\cite{MusicLM_paper} & High-fidelity music generation from text prompts & Not released & SoundStream, w2vBERT, and MuLan & Prompt-based \\
\hline
CMT\cite{Di_2021} & Video background music generation & Python & Transformer & Visual-based (video) \\
\hline
V-MusProd\cite{zhuo2022video} & Music generation from video with decoupling chords, melody, and accompaniment & Not released & Custom & Visual-based (video) \\
\hline
Foley Music\cite{foley_music} & Music generation from video clips using foley sounds & PyTorch & Encoder-decoder & Visual-based (video) \\
\hline
Commercial Tools & Music generation with customizable parameters &Web User Interface & Not specified & Parameter-based and prompt-based \\
\hline
\end{tabular}
\label{table2}
\end{table}

\subsubsection{Non-Neural Network Approaches}\label{non-neural-network-methods}
Based on our studies and the assistance of large language models (Bard \footnote{https://bard.google.com}, ChatGPT 3.5 \footnote{https://openai.com/blog/chatgpt}, Claude1 \footnote{https://www.anthropic.com/product}) we identified there are three non-neural network approaches to construct music; including, Markov Chains, Rule-based models, and Evolutionary algorithms. Table \ref{table1} illustrates these methods, their key functionalities, and their underlying models. The Classification column in our tables, as discussed in Section \nameref{sec:categorization}, categorizes the music generation tools based on the type of input they require to generate music.

Markov chains are mathematical models that are used to analyze and predict the behavior of systems that exhibit a property called the Markov property  \cite{GUDIVADA2015203}. This property states that the future state of a system depends only on its current state, and not on its past history. By predicting the probability of transitions between different musical notes or events, they have been used in melody generation to create new melodies that flow smoothly and naturally \cite{MarkovProcess}. Later, Markov chains have also been used to create customized music based on user manual input of moods \cite{Ramanto2017MarkovCB}. Recently, Shapiro et al. \cite{Shapiro2021} employ Markov chains for music generation to analyze patterns in existing music and create new compositions. 

Another group of works focuses on Rule-based music generation, which employs predefined rules to create musical data that follows specific patterns or styles. An instance of this approach is provided by Spangler \cite{rule-based-generation}, who introduced a system for real-time accompaniment. This system extracts and employs a set of harmonic rules from musical examples to generate new harmonies in response to an input melody, demonstrating the application of deterministic rule sets in music generation. On a different note, Sneyers and De Schreye \cite{APOPCALEAPS} developed APOPCALEAPS, a system for generating and learning pop music that leverages a unique form of probabilistic rules. These rules, learned automatically from examples, allow for a range of potential musical outcomes, providing an element of randomness within defined parameters. This chance-based methodology has been shown to foster versatility and creativity in music generation.

A third group of works focused on evolutionary algorithms, specifically genetic algorithms, in the sphere of music generation. These algorithms engage in the selective identification of the best musical sequences, refining them through mutation and crossover, thus fostering the genesis of novel compositions. In 1994, Biles introduced GenJam (Genetic Jammer) \cite{Biles1994}, an interactive genetic algorithm that evolves jazz solos improvisations by evolving music measures populations, with the algorithm using real-time feedback from a human mentor as a fitness function. Later in 2000, Marques et al. \cite{marques2000music} applied genetic and evolutionary algorithms to music generation, wherein melodic and musical theory concepts were used as a fitness function, enabling efficient navigation through large search spaces to find creative solutions. More recently, in 2021, Rocha de Azevedo Santos et al \cite{shu28011} employed the genetic algorithm to create a piano music composition system using mood templates. This system evolves random music pieces until their mood aligns with the template, using MIDI statistical features to calculate the music-template distance. Despite some constraints, such as the computer-generated feel due to the lack of rhythm regularity and synchronicity, these examples underscore the potential of evolutionary algorithms in the autonomous generation of music and their role in emulating modern composers' creative strategies.

\subsubsection{Neural Network-based Music Generation}
In the following section, we undertake a systematic examination of music generation models driven by neural networks. The discourse is partitioned into three distinct subsections, each focused on a particular type of model: parameter-based models, prompt-based models, and visual-based models. In each case, we provide a thorough examination of the model architecture, delineating its key features and capabilities. Additionally, we offer a balanced assessment of the benefits and potential limitations of each model. Our objective is to provide a holistic view of the landscape of neural network-based music generation models, elucidating the various mechanisms by which they operate and their respective merits and drawbacks.

Table \ref{table2} provides a summary of various music generation tools based on their main function, interface access, neural network models, and classification. Note that some models were not accessible due to restrictions implemented by the owner corporation.  

\hypertarget{parameter-based-music-generation-tools}{%
\section{Parameter-based Music Generation
Tools}\label{parameter-based-music-generation-tools}} Parameter-based Music Generation is a specific category of models requiring certain input parameters to function effectively. The parameters in question could range from musical attributes like 'tempo' or 'key' to more abstract inputs such as 'mood' or 'temperature'.

Unlike their Parameter-free counterparts, which function autonomously without the need for specific inputs, Parameter-based models offer a higher degree of control to the users, allowing them to steer the generated music in a direction that aligns with their preference. This is achieved by tailoring the input parameters which guide the music generation process.

\subsection{Magenta} 

\subsubsection{Overview} Magenta\cite{google_ai}, an open-source research initiative, is as an expansive  framework exploring the incorporation of machine learning in the artistic and musical creative processes. The authors described that their focus is to explore the possibility of developing intelligent tools and interfaces that facilitate the integration of music generation models into the creative process of artists and musicians. Their goal is not to replace the creative process but to enhance it.

Magenta's primary repositories, including its popular Python TensorFlow library, can be found on GitHub\footnote{\url{https://github.com/magenta}}, offering an array of neural network models. These encompass multiple Sequential (RNN/LSTM\cite{RNN_LSTM, Sherstinsky_2020, LSTM_intro}) models tailored for generating drum and melody sequences, the MusicVAE\cite{roberts2019hierarchical}, and GANSynth\cite{engel2018gansynth} which utilizes generative adversarial networks\cite{NIPS2014_5ca3e9b1} for audio synthesis.

In this paper, we present an examination of several music generation models created by Magenta, specifically, Music VAE, NSynth, Melody RNN, and Drums RNN. These models were selected due to their recognized influence within the field and their relevance to contemporary trends in music generation. This estimation of popularity was made through an indirect assessment using large language models, such as ChatGPT, to ascertain the frequency and context of their mentions within relevant literature and digital discourse.

\subsubsection{Architecture} Magenta contains several neural network  models for music generation, which can be classified into three distinct model types:
sequential models \cite{RNN_LSTM} such as recurrent neural networks (RNNs) and Long Short-Term Memory network, variational autoencoders (VAEs) \cite{roberts2019hierarchical}, and
neural synthesizer (NSynth)\cite{engel2017neural}. We will explain each model briefly in this section.

Sequence-based music generation models, such as Melody RNN \cite{magentaMelodyRNN}, Improv RNN \cite{magentaImprovRNN}, and
Polyphony RNN \cite{magentaPolyphonyRNN}, are trained to learn the distribution of musical patterns and structures in a given dataset. These models are capable of
generating new music by predicting the next note in a sequence of musical notes given the previous ones. The learned distribution of musical patterns and structures from training data enables the models to generate music in a variety of styles and genres.

VAEs are probabilistic generative models that learn the probability distribution of the input dataset \cite{kingma2022autoencoding}. Music VAE \cite{roberts2019hierarchical}, a hierarchical recurrent VAE for music generation, and it is capable of generating new music by sampling from the learned distribution. The model is trained on a dataset of musical compositions of approximately 1.5 million MIDI files collected by the author\cite{roberts2019hierarchical}, and it learns a low-dimensional representation of the input data, which can be used to generate new music with various styles
and characteristics. The reconstruction quality was also evaluated on the publicly available Lakh MIDI Dataset \cite{Raffel2016-et}.

NSynth\cite{engel2017neural} is a neural autoregressive model for music generation \cite{dubreuil2020hands}, uses a WaveNet \cite{oord2016wavenet} and an autoencoder to generate high-quality music with complex and diverse sound characteristics. Autoregressive models learn the probability distribution of the next note in a sequence given the previous ones. However, they require a large amount of computational resources and a large dataset of musical examples for training.

\subsubsection{Features and capabilities} The core foundation of Magenta revolves around the concept of music note sequences, which are an abstract representation of a
series of notes with varying pitches, instruments, and strike velocities, similar to MIDI. As demonstrated on the official website \cite{google_ai_start}, the primary usage of Magenta involves generating MIDI-style music with a single-track melody.

NoteSequence is a data structure that captures various aspects of music, such as timing, pitch, and instrument information. The NoteSequence library is an essential component of the Magenta project, and it focuses on the serializable NoteSequence representation and provides a variety of utilities for working with musical data. The main functionalities of the library include creating note sequences from different formats (e.g., MIDI, abc\cite{abcnotation},
  MusicXML\cite{musicxml}), manipulating note sequences (e.g., slicing, quantizing), extracting components from note sequences (e.g., melodies, drum tracks, chords), exporting note sequences to different formats (e.g., MIDI or audio) and converting note sequences to representations useful for model training
  (e.g., one-hot tensors).

A summary of the functionalities of Magenta models is presented in Table \ref{table3}. 

\begin{table}[htbp]
\centering
\caption{List of the Magenta Models in Model Directory\cite{magenta-models}}
\renewcommand{\arraystretch}{1.5}
\small
\begin{tabular}{|p{5cm}|p{10cm}|}
\hline
\textbf{Model} & \textbf{Description} \\
\hline
Arbitrary Image Stylization & A machine learning system for performing fast artistic style transfer that may work on arbitrary painting styles. \\
\hline
Coconet & Trains a CNN to complete partial musical scores. \\
\hline
Drums RNN & Applies language modeling to drum track generation using a LSTM. \\
\hline
GANSynth & Uses a GAN-based model to generate audio\cite{engel2017neural}. \\
\hline
Image Stylization & A "Multistyle Pastiche Generator" that generates artistic representations of photographs. Described in "A Learned Representation For Artistic Style". \\
\hline
Improv RNN & Generates melodies conditioned on an underlying chord progression using an LSTM. \\
\hline
Melody RNN & Applies language modeling to melody generation using an LSTM. \\
\hline
Music Transformer & Generates musical performances, either unconditioned or conditioned on a musical score. \\
\hline
Music VAE & A hierarchical recurrent variational autoencoder for music. \\
\hline
NSynth & Neural Audio Synthesis with WaveNet Autoencoders". \\
\hline
Onsets and Frames & Automatic piano music transcription model with a dual-objective approach. \\
\hline
Performance RNN & Applies language modeling to polyphonic music using a combination of note on/off, timeshift, and velocity change events. \\
\hline
Piano Genie & Learns a low-dimensional discrete representation of piano music using an encoder-decoder RNN. \\
\hline
Pianoroll RNN-NADE & Applies language modeling to polyphonic music generation using an LSTM combined with a NADE. \\
\hline
Polyphony RNN & Applies language modeling to polyphonic music generation using an LSTM. \\
\hline
RL Tuner & Enhances an LSTM trained for monophonic melody generation using reinforcement learning (augmenting deep Q-learning with a cross-entropy reward). \\
\hline
Score2Perf & Generates musical performances, either unconditioned or conditioned on a musical score. \\
\hline

\end{tabular}
\label{table3}
\end{table}

In the following, we explain Magenta's different neural network models  to achieve various tasks and their respective functions.

Nsynth: Neural Audio Synthesis (NSynth)  model uses WaveNet-based autoencoder to generate audio\cite{engel2017neural}. Unlike traditional synthesizers that generate audio from hand-designed components, such as oscillators and wavetables, NSynth uses deep neural networks to generate sounds at the level of individual samples. By learning directly from data, NSynth provides musicians with intuitive control over timbre and dynamics, allowing them to explore new sounds that would be difficult or impossible to produce with a hand-tuned synthesizer.

One of the key features of NSynth is its ability to factorize the generation of music into notes and other musical qualities, such as timbre. It assumes that the generation of music can be modeled by factoring the quality of the notes and the other musical qualities. To facilitate this, the developers built the NSynth dataset \footnote{https://magenta.tensorflow.org/datasets/nsynth}, which is a large collection of annotated musical notes sampled from individual instruments across a range of pitches and velocities. Alongside the dataset, the developers also released the WaveNet-style autoencoder model, which learns codes that meaningfully represent the space of instrument sounds.

Melody-rnn: Melody RNN has four configurations: basic, mono, lookback, and attention. The basic configuration uses one-hot encoding to represent extracted melodies as input to the LSTM. For training, all examples are transposed to the MIDI pitch range {[}48, 84{]}, and outputs are also in
this range. The mono configuration is similar to the basic configuration but can use the full 128 MIDI pitches. The lookback configuration introduces custom inputs and labels that allow the model to recognize
patterns that occur across one and two bars and helps the model recognize patterns related to an event's position within the measure. Finally, the attention configuration uses attention mechanisms to access past information without having to store that information in the RNN cell's
state. This allows the model tolearn longer-term dependencies and generates melodies with longer arching themes.

To optimize the performance of the Melody RNN model, several hyperparameters can be adjusted. These hyperparameters include the batch size, which determines the number of sequences used in each training iteration, as well as the number of time steps in the generated sequence, learning rate, and the number of hidden units in each LSTM layer. Furthermore, the dropout keep probability is used to determine the probability that a given hidden unit will be retained during training, and the attention length specifies the length of the attention mechanism. Additionally, the number of training steps can be specified to determine when to stop the training loop, while the hyperparameter eval ratio is used to determine the fraction of examples reserved for evaluation during training. By tuning these hyperparameters, users can customize the model's performance to generate melodies that better suit their needs.

Music VAE: MusicVAE is another Magenta model based on Virtual Auto Encoder (VAE) \cite{kingma2022autoencoding}, provides
different modes of interactive musical creation, including random sampling from the prior distribution, interpolation between existing sequences and manipulation of existing sequences via attribute vectors or a latent constraint model. Representations of melodies/bass-lines and drums are based on those used by MelodyRNN and DrumsRNN (described in the following sections).

MusicVAE uses pre-trained checkpoints for different configurations, which are used to generate outputs from the command line.

GrooVAE is a variant of MusicVAE for generating and controlling expressive drum performances. It can be trained with Groove MIDI Dataset \cite{groove2019} of drum performances.

Drums RNN: This model generates drum tracks. While modeling drum tracks, it is considered a single sequence of events by mapping all of the different MIDI drums onto a smaller number of drum classes and representing each event as a single value. This value presents the set of drum classes that are struck. Unlike melodies, drum tracks are polyphonic, and multiple drums can be struck simultaneously.

The model provides two setups, one\_drum and drum\_kit, where the former encodes all drums to a single class, while the latter incorporates a nine-piece drum kit, including different drum types encoded as a 512 length one-hot vector, and binary counters augment the model's input. Drums RNN can operate either with a pre-trained model, where users input the necessary parameters and potentially specify primer drums, or train a fresh model by converting MIDI files into NoteSequences, which transform into Sequence examples for training and evaluation. 

\subsubsection{Advantages and Limitations}
One main advantage of Magenta is that it is an open-source research project that aims to explore the role of machine learning as a tool in the creative process of art and music. It has developed a vast range of
models and libraries since 2016 that provide various functions for creating, manipulating, and generating music and art. \\
Magenta provides a wide range of resources and libraries (listed in \ref{table2}), which allows
for flexibility in the input and manipulation of music and art data. In addition, each model has its unique approach and can generate different types of music.
This provides a wealth of possibilities and functionalities for users to
explore and utilize\footnote{As of April 2023, the Magenta Python library has 3,700+ forks on GitHub, while Magenta.js has 300+ forks, demonstrating widespread adoption. 
}. 
In addition, Magenta supports multiple formats, such as MIDI and MusicXML, which allows for interoperability with other music and art software. \\
However, Magenta also has some limitations. It is a complex project that requires a certain level of technical knowledge and expertise to
use and understand. Some models and libraries may be geared toward specific use cases, limiting their overall versatility. Musicians may face challenges in learning and selecting the right tools due to the extensive range of models and libraries available.   It should be noted that Magenta's main output is in the form of MIDI-style music sequences that may only be readily usable by individuals with prior knowledge of music production.  Magenta cannot always produce the desired result, and the generated music may require further manual fine-tuning and adjustments afterward.

\subsection{JukeBox} 

\subsubsection{Overview} Jukebox \cite{jukebox,  openai_2020} is another neural network based model, for music generation. It includes vocal elements in a diverse range of musical genres
and styles. The output of the Jukebox model is presented as raw audio. This model utilizes VQ-VAE and Transformers to produce original
compositions. It is noteworthy that Jukebox is capable of generating singing, which is a unique feature compared to other music generation models at the time of its release. 

\subsubsection{Architecture} Jukebox uses hierarchical VQ-VAE \cite{jukebox} architecture to compress music into discrete codes. This method is similar to the VQ-VAE-2 \cite{razavi2019generating} model used for image generation, but it has been modified to suit the needs of music generation better. The model uses random restarts to prevent codebook collapse, which is a common issue for VQ-VAEs. It separates decoders to maximize the use of bottlenecked top levels in hierarchical VQ-VAE, and a spectral loss to allow for the reconstruction of higher frequencies. The model has three levels, each compressing the audio by a different factor and reducing the detail in the audio. However, it is still able to retain important information about the pitch, timbre, and volume of the audio. 

Following this compression, a set of prior models, including a top-level prior and two upsampling priors, are trained to learn the distribution of music codes in the compressed space. These prior models use a simplified variant of Sparse Transformers, a form of self-attention, for efficient training. Serving as probabilistic representations of the underlying structure in the compressed music space, the prior models enable Jukebox to generate new music samples by capturing global patterns and finer-grained details through a hierarchical approach.

To train the model, a dataset of 1.2 million songs was collected, along with corresponding lyrics and metadata. Additionally, the model can be conditioned on the artist and genre, which allows it to generate music in a specific style. Artists and genres can be clustered together using t-SNE \cite{van2008visualizing}, which reflects the similarities and reveals unexpected associations between them.

\subsubsection{Features and Capabilities} Jukebox features two types of models; one with lyrics and one without lyrics. The models are typically named with a prefix that indicates the number of parameter upsamplers, such as "5b" or "5b\_lyrics". This naming convention highlights the upgrade from the previous "1b" version, with the numerical value representing the increase in parameter upsamplers for enhanced performance.

It has several hyperparameters, and we list a few of the important ones. One of the hyperparameters is \textit{speed\_upsampling}, which allows for faster upsampling but may result in slightly "choppy" samples. Another hyperparameter is the \textit{mode}, which can be either "primed" or "ancestral". The primed mode continues an existing song, while the ancestral mode generates a song from scratch. If the primed mode is selected, an audio file is required to specify the song Jukebox will continue.

Jukebox also provides the ability to select the artist and genre of the music to be generated. The available options for artists and genres can be found in the GitHub repository for \footnote{https://github.com/openai/jukebox}. The 5b\_lyrics and 5b models use the v2 genre lists, while the 1b\_lyrics model uses the v3 genre lists. Up to five v2 genres can be combined, but combining v3 genres is impossible.

Finally, Jukebox provides a hyperparameter called \textit{sampling\_temperature},
which determines the creativity and energy of the generated music. The higher the temperature, the more chaotic and intense the result. Keeping the temperature between .96 and .999 is recommended, but users can experiment with different values to achieve the desired results.

\subsubsection{Advantages and Limitations }The Jukebox model has several advantages that make it a powerful tool for music generation. One major advantage is its use of the VQ-VAE
technique, which allows for the efficient compression of music into discrete codes. This allows for the generation of novel songs while retaining important information about the pitch, timbre, and volume.

The Jukebox model also has the ability to be conditioned on artist and genre, which allows it to generate music in a specific style. This allows for more control over the generated music and can be useful for applications such as creating custom soundtracks or personalized playlists.

Another advantage of the Jukebox model is its use of a dataset of 1.2 million songs for training, paired with corresponding lyrics and metadata, which allows it to learn a wide variety of musical styles and patterns.

However, there are also some limitations to the Jukebox model. One limitation is that the downsampling process used to compress the audio can result in a loss of detail in the generated music. Additionally, the model's ability to generate music in a specific style is limited by the diversity of the dataset that it was trained on. If a specific style of music is not well represented in the dataset, the model may struggle to generate music in that style. 

While Jukebox shares the common limitation of data-driven music generation models in producing variations of existing patterns, its unique challenge is the computational complexity involved in its hierarchical approach. The generating process might take a significant amount of resources for commonly accessible machines. This can be a barrier for developers and researchers who do not have access to large corporation GPU clusters.

\subsection{MuseNet}

\subsubsection{Overview} MuseNet\cite{openai-musenet} is a deep neural network designed with the ability to create 4-minute musical compositions with up to 10 different instruments and has the potential to blend various styles.
\subsubsection{Architecture} MuseNet\cite{openai-musenet} utilizes a transformer model \cite{gpt-2, vaswani2017attention}, similar to GPT-2, predicting subsequent tokens in a sequence, applicable to both audio and text. Using the optimized kernels of Sparse Transformer \cite{sparse-transformer}, a 72-layer network with 24 attention heads is constructed, focusing on a context comprising 4096 tokens.\\
The model's large context window may contribute to its ability to capture and maintain long-term structural elements in music, crucial in generating coherent musical pieces\cite{openai-musenet}.
Sequential data is used for training, with the aim to predict the subsequent note(s) given a set of notes \cite{openai-musenet}. The dataset used for training was sourced from various collections including ClassicalArchives \cite{classicalarchives}, BitMidi \cite{bitmidi}, and the MAESTRO dataset \cite{maestro-dataset}.\\
Various encoding strategies were used to convert MIDI files into a model-friendly format. An encoding scheme combining pitch, volume, and instrument information into a single token was finally adopted, balancing expressivity and conciseness\cite{openai-musenet}.\\
Further, the architecture incorporates several embeddings for structural context, including positional embeddings, time-tracking embeddings, and chord-specific embeddings\cite{openai-musenet}.\\
During training, data augmentation techniques such as transposition, volume and timing augmentation, and mixup in the token embedding space were used \cite{mixup}. Additionally, an "inner critic" mechanism is employed, training the model to discern between original dataset samples and its own past generations \cite{openai-musenet}.

\subsubsection{Features and Capabilities} MuseNet offers users the opportunity to explore and interact with its capabilities through two modes: "simple" and "advanced". The simple mode provides users with a pre-generated set of uncurated samples, which allows them to experiment with different composers, styles, and starting points. This feature enables users to explore the diverse range of musical styles the model can generate. On the other hand, the advanced mode offers a more interactive experience where users can directly interact with the model to create entirely new pieces. Although this mode may require longer completion times, it offers the ability to generate a unique and original piece. By offering these two modes, MuseNet allows users to experiment with its capabilities and makes it accessible to both novice and experienced users.\\
To create more controlled generations, composer and instrumentation tokens were introduced during the training phase. These tokens provide contextual cues to the model about the expected style and instrumentation of the music, enabling users to guide the model's output in a desired direction \cite{openai-musenet}.

\subsubsection{Advantages and Limitations } MuseNet exhibits a remarkable ability to retain the long-term structure of a musical piece, thereby effectively imitating the style or musician with precision. This proficiency is achieved through its architecture, which has been meticulously engineered to generate musically coherent and aesthetically pleasing compositions when parameters congruent with its training and encoding scheme are utilized. These parameters comprise appropriate prompts like composer and instrumentation tokens and a consolidated encoding of musical features such as pitch, volume, and instrument information into a single token. The precise amalgamation of these components endows the model with the capability to produce expressive yet concise musical sequences, effectively simulating natural compositions.\\
However, there are limitations to be considered when using MuseNet. Although the user can suggest specific instruments, MuseNet generates each note by calculating the probabilities across all possible notes and instruments. Therefore, the model may choose different instruments from what the user suggested. Moreover, MuseNet needs help with odd pairings of styles and instruments, such as Chopin with bass and drums. Therefore generations will be more natural if the instruments closest to the composer or band's usual style are selected. Lastly, MuseNet does not guarantee that the music generated is free from external copyright claims.

\subsection{Music Transformer} 

\subsubsection{Overview} Music Transformer is another neural network model designed to generate long sequences of music with long-term structure\cite{huang2018music}. It uses the transformer architecture and is capable of scaling to musical sequences on the order of minutes.
\subsubsection{Architecture} The Music Transformer model uses discrete tokens to represent music, with the specific vocabulary determined by the training dataset. The Transformer decoder is a generative model, relying on self-attention mechanisms with learned or sinusoidal position information. Each layer consists of a self-attention sub-layer followed by a feedforward sub-layer. The attention layer uses the scaled dot-product attention mechanism, and the feedforward sub-layer performs two point-wise dense layers. The model employs relative position representations to allow attention to be informed by the distance between positions in a sequence. The Music Transformer reduces the intermediate memory requirement of close attention through a "skewing" procedure\cite{huang2018music}. Also, it uses relative local attention for very long sequences by chunking the input sequence into non-overlapping blocks.
\subsubsection{Features and Capabilities} Music Transformer is available as both TensorFlow \footnote{https://github.com/jason9693/MusicTransformer-tensorflow2.0} and PyTorch \footnote{https://github.com/jason9693/musictransformer-pytorch} versions with easy-to-use APIs. It can generate music with long-term structure and scales to sequences of music on the order of minutes. In addition, it has a reduced memory footprint, allowing it to generate longer sequences of music. The model is trained on sequential data and can generate natural compositions when the correct hyper-parameters are selected.
\subsubsection{Advantages and limitations } Music Transformer has the advantage of generating long music sequences with long-term structure and has been modified to reduce memory requirements. However, the melody produced by the tool can be monotonous if underfitting or insufficient training data exists. This could limit the tool's usefulness for complex music generation tasks and may require additional optimization or training with larger and more diverse datasets to improve the quality of the generated music.

\subsection{SDMuse}
\subsubsection{Overview} SDMuse is a unified stochastic differential music editing and generation framework that can generate and modify existing musical pieces at a fine
granularity proposed in 2022\cite{sdmuse, zhang_ren_zhang_yan_2022}.
\subsubsection{Architecture } The framework follows a two-stage pipeline with a hybrid representation of pianoroll and MIDI-event. In the first stage, SDMuse generates and edits pianoroll through an iterative denoising process based on a stochastic differential equation (SDE) using a diffusion
model generative prior\cite{song2021scorebased}. In the second stage, the framework refines the
generated pianoroll and predicts MIDI-event tokens auto-regressively.
\subsubsection{Features and Capabilities} SDMuse can compose a whole musical piece from scratch and modify existing musical pieces in various ways, such as combinations,
continuation, inpainting, and style transfer. The framework can also extract fine-grained control signals from the musical piece itself without requiring extra data annotation, making it more efficient.\\
The proposed framework takes advantage of the two most common music representations, pianoroll, and MIDI-event. Pianoroll-based methods use pianorolls to represent music scores, while MIDI-event-based methods convert musical pieces to a MIDI-event token sequence. The hybrid representation used in SDMuse is more appropriate for extracting and
controlling perceptive information like structure while generating and modeling precise music performance details, such as velocity and fine-grained onset position.\\
SDMuse involves several fine-grained control signals, including note density, pitch distribution, and chord progression sequence during the training process of the diffusion model to enable unconditional and conditional music generation/editing at the same time. These control signals can be extracted from the musical piece itself, making it more efficient.
\subsubsection{Advantages and Limitations } Although we could not experiment with SDMuse practically due to the lack of a demonstrative version available for use, the model's capabilities in generating
and editing multi-track music have been demonstrated in the academic paper. Specifically, the model can perform fine-grained editing using stroke-based generation and editing, inpainting, outpainting, combination, and style transfer. Furthermore, the effectiveness of SDMuse in generating and editing music has been evaluated in various experiments using the ailabs1k7 pop music dataset \cite{hsiao2021compound}. Despite the lack of first-hand experimentation, the reported results and methodology provide insight into the potential of SDMuse as a music generation and editing tool. \\
Overall, SDMuse's effectiveness in generating and editing music pieces has been demonstrated through experiments on various music editing and generation tasks using the ailabs1k7 pop music dataset.

\subsection{Music Generation with Sentiment based on mLSTM}

\subsubsection{Overview } The Music Generation with Sentiment \cite{ferreira2021learning} is a model that can compose music with a specified sentiment. It is based on mLSTM, which was initially designed for generating Amazon product reviews based on positive or negative emotions \cite{radford2017learning}. In this model, a music piece is represented as a sequence of words and punctuation marks from a vocabulary that represents events retrieved from the MIDI file. The model can also be used for sentiment analysis of symbolic music, enabling users to categorize music based on emotional content.
\subsubsection{Architecture } The model is trained on a dataset called VGMIDI, which is composed of 823 pieces extracted from video game soundtracks in MIDI format. The pieces are piano arrangements of the soundtracks and vary in length from 26 seconds to 3 minutes. Among these pieces, 95 are annotated according to a 2-dimensional model that represents emotion using a valence-arousal pair. Valence indicates positive versus negative emotion, and arousal indicates emotional intensity \cite{russell1980}. The valence-arousal model is also one of the most common dimensional models used to label emotion in music. \\
It uses mLSTM and logistic regression to compose music with a sentiment. The sentiment in music is determined by various characteristics such as melody, harmony, tempo, and timbre. Additionally, the same labeled data can be used to explore affective algorithmic music composition in both classification (multiclass and/or binary) and regression problems.
\subsubsection{Features and Capabilities} The model generates music that aligns with a specific mood or emotion, providing users with a convenient tool for generating music with a desired sentiment. Moreover, the model can be utilized for sentiment analysis of symbolic music, facilitating users to examine and classify music based on emotional content.
\subsubsection{Advantages and Limitations} According to the results in the paper, the generative mLSTM model combined with logistic regression demonstrated substantial classification accuracy (89.83\%); the model achieved 84\% accuracy for generating positive music pieces and 67\% accuracy for negative ones. However, the negative pieces were found to be more ambiguous, indicating room for further improvement. The developers acknowledge this as a limitation of the model and plan to improve its ability to generate less ambiguous negative pieces in future work. Additionally, authors propose that their future work expand the model's capability to generate music with specific emotions (e.g., happy, sad, suspenseful) and valence-arousal pairs (real numbers), and also compose soundtracks in real-time for oral storytelling experiences.

\section{Prompt-Based Music Generation Tools}
Prompt-based music generation converts textual inputs into music. These tools utilize machine learning and language processing algorithms to interpret text prompts, transforming the embedded semantics and sentiments into harmonious musical compositions.

\subsection{Riffusion} 

\subsubsection{Overview}  Riffusion\cite{Forsgren_Martiros_2022} is a fine-tuned diffusion model that generates audio clips from text-based prompts and images of spectrograms. The model
architecture is based on the Stable Diffusion model, an open-source AI
model that generates images from text.
\subsubsection{Architecture} The architecture of Riffusion is primarily built upon the Stable Diffusion v1.5 model\footnote{https://huggingface.co/runwayml/stable-diffusion-v1-5}, a deep learning architecture that excels in generative tasks. In essence, this model uses a variant of denoising autoencoders combined with a diffusion process to model the data distribution. This architecture has been fine-tuned for the unique task of generating audio from spectrogram images.

The backbone of the model's input is the spectrograms that represent sound in a visual form. These spectrograms are processed by a neural network architecture designed to understand and generate complex images. Given the nature of spectrograms, with their intricate structures representing different audio frequencies over time, a network that can handle such data complexity is crucial. Hence, the architecture of Stable Diffusion, which has been proven to generate intricate image structures, becomes an apt choice.

The model architecture is equipped with Image-to-Image transformation capability, a critical feature that allows the model to condition its output not only based on a text prompt but also on other images. This feature effectively enables the architecture to adapt to the structural nuances of different input images, driving its ability to generate a variety of audio outputs. In addition, the architecture also includes Looping and Interpolation mechanisms, which require the model to navigate smoothly through the latent space, thus preserving the coherence in audio output. 
\subsubsection{Features and Capabilities } Riffusion provides a web-based user interface that enables users to select a prompt or input customized text containing keywords, such as
"jazz" or "guitar," and generates beats based on these keywords. The platform is capable of generating an infinite number of prompt variations based on different seeds. Additionally, Riffusion can generate both short music and loopable jams, and model checkpoints are provided for user convenience. \\
The Riffusion web-based user interface \footnote{https://www.riffusion.com/} provides an accessible and user-friendly approach for generating beats. Users can adjust parameters such as seed image and denoising level to vary the music they generate and save the music as short riffs.
\subsubsection{Advantages and Limitations} Riffusion, as a music generating tool based on diffusion models, boasts several advantages. Firstly, it has a user-friendly interface that allows users to easily generate music from text input without the need for coding or complicated installation procedures. Additionally, Riffusion produces high-quality music with minimal noise, making it an excellent option for those seeking high-quality music output.\\
However, the inherent architecture of the Riffusion platform results in certain limitations in user control over the musical output. Specifically, the music generated is primarily based on the provided text input and seed images, which guide the stable diffusion model in generating the output audio. This reliance on text prompts and seed images means that the user's ability to control the specifics of the musical output is contingent upon the quality and relevance of these inputs to the desired output. Furthermore, the number of seed images available for generating music is limited, which may result in a lack of variety in the music created. Another area for improvement of Riffusion is its inability to download longer compositions, with users only able to download short tracks. Additionally, the looping feature of the short jams may result in repetitiveness. \\
While Riffusion's lack of flexibility may not suit users who require a high degree of customization, it is essential to note that this design choice was made to create an accessible and straightforward music generation tool that can quickly create music based on user input, without requiring extensive musical knowledge or technical skills.

\subsection{Noise2Music} 

\subsubsection{Overview} Noise2Music  explores the utilization of diffusion models in generating high-quality music audio from text prompts\cite{huang2023noise2music}. Similar to MusicLM, the model is also based on MuLan\cite{huang2022mulan}, which is a joint embedding model that connects music to unconstrained natural language
music descriptions.
\subsubsection{Architecture} The End-to-End Diffusion architecture utilizes a series of diffusion models in succession to generate the final music clip. This process
involves the use of pseudo-labeling for training data generation, where a large unlabeled set of music audio is used to train two deep models. A large language model is employed to generate a vast collection of generic music descriptive sentences as caption candidates, while a pre-trained music-text joint embedding model is utilized to assign the captions to each music clip via zero-shot classification \cite{palatucci2009zero}. This method generates the training set by pseudo-labeling, enabling the model to handle complex, fine-grained semantics beyond simple music label conditioning. The Noise2Music models have been shown to have strong generative ability, capable of producing music pieces that go beyond basic music label conditioning, utilizing the semantic content of the captions provided to generate complex and detailed music.
\subsubsection{Features and Capabilities} Noise2Music is a music generation model that demonstrates generative ability grounded on semantically rich text prompts. It can generate music based on a range of key musical attributes, including genre, instrument, tempo, mood, vocal traits, and era. Additionally, it can generate music based on creative prompts.
\subsubsection{Advantages and Limitations} Noise2Music has advantages in providing high-quality music from rich text prompts, offering creativity for artists and content
creators. However, there are limitations to the model, such as potential biases learned from the training sets, which can manifest in subtle and unpredictable ways. Furthermore, misappropriation is also a risk when the created content matches the training data's examples. Therefore, responsible model development practices, duplication checks, and efforts to identify and address potential safety issues are necessary for improving these generative models. The model was not released due to the limitations and risks.

\subsection{Moûsai} 
\subsubsection{Overview } 
Moûsai \cite{schneider2023mousai} is a text-to-music generation tool based on latent diffusion model \cite{rombach2022highresolution}. 

\subsubsection{Architecture } 
Moûsai employs a two-stage cascading diffusion methodology to generate high-quality stereo music from textual descriptions. The system generates music at 48kHz, lasting multiple minutes.

The first stage uses a Diffusion Magnitude Autoencoder (DMAE) \cite{DMAE} to compress the audio waveform by a factor of 64. This significant reduction in data size does not compromise the high fidelity of the audio reproduction. The DMAE conditions the diffusion process on a compressed latent vector of the input, thereby creating a powerful generative decoder.

The second stage of Moûsai generates a novel latent space using the diffusion model, conditioned on text embeddings. These embeddings are produced from a given description using a frozen transformer language model. Moûsai's two stages employ an efficient one dimensional U-Net architecture with different configurations. The latent text-to-audio diffusion stage applies diffusion to the compressed space created in the first stage and uses cross-attention blocks to provide the conditioning text embedding.

Moûsai was trained on a diverse dataset of 2,500 hours of stereo music, sampled at 48kHz. The metadata such as title, author, album, genre, and year of release were used as the textual description. To enhance the robustness of the conditioning, each element of the metadata was dropped with a probability of 0.1. Training could be completed within a week using an A100 GPU with a batch size of 32. During inference, a novel audio source of about 88 seconds can be synthesized in less than 88 seconds using a consumer GPU.

\subsubsection{Features and Capabilities }
The standout feature of Moûsai is the ability to generate multiple minutes of high-quality stereo music at a 48kHz resolution purely from textual descriptions. This capability serves to bridge the gap between textual ideas and musical creation, providing a rich platform for translating imaginative concepts into audibly engaging music.

The input to the model can include not only descriptive text but also musical features such as temporal dimension, long-term structure, overlapping sound layers, and subtle nuances. This holistic approach to input parsing allows for a comprehensive understanding of the musical intention, resulting in high fidelity musical output.

Training on an extensive and diverse dataset of 2,500 hours of stereo music samples equips Moûsai with a broad 'vocabulary' of musical elements. This, coupled with the capacity to create a diverse range of music, elevates the versatility and adaptability of the model.

Open source samples of music generated using Moûsai are provided as a testament to its capabilities \footnote{URL: \url{https://anonymous0.notion.site/anonymous0/Mo-sai-Text-to-Audio-with-Long-Context-Latent-Diffusion-b43dbc71caf94b5898f9e8de714ab5dc}}.

\subsubsection{Advantages and Limitations } 
Moûsai's key advantage lies in its capacity to generate long high-quality stereo music from textual descriptions. This ability to translate linguistic semantics into musical language opens up vast possibilities for musical creativity and innovation. 

However, the model's complexity and the demand for substantial computational resources might present challenges for users with limited processing power. Furthermore, the interpretability of Moûsai could be a hurdle, given the complexity and the non-intuitive nature of the diffusion processes it uses to generate music.

The model's proficiency in generating specific music genres might also be limited due to the distribution of its training data. While it has been exposed to a wide range of genres such as electronic, pop, metal, and hip hop, it has less experience with others like blues and classical. As a result, the quality of generated music in these less exposed genres might not be as robust.

\subsection{MusicLM}

\subsubsection{Overview } MusicLM is a recently developed music generation model \cite{MusicLM_paper} that is capable of generating high-fidelity music from rich
text descriptions. The model has gained attention due to its ability to
produce music that can last for minutes and its high audio quality
performance. Lamentably,  MusciLM model has not been officially released as an
open-source project, the team has provided demo audios and the MusicCaps
dataset that contains 5.5k music-text pairs. However, some open-source implementation of MusicLM architecture is proposed such as:  \cite{bouchard_musiclm-pytorch_2023}.
\subsubsection{Architecture } The MusicLM model by Agostinelli et al. \cite{MusicLM_paper} is composed of three pre-trained models, SoundStream\cite{zeghidour2021soundstream}, w2vBERT\cite{chung2021w2vbert}, and MuLan\cite{Mulan2022}. These models are used to extract audio representations for use in text-conditioned music generation. In particular, SoundStream is used to extract self-supervised audio representations from monophonic audio data. These acoustic tokens are used for high-fidelity synthesis. w2vBERT is used to extract semantic tokens from the audio data, which facilitates long-term coherent generation. MuLan is used to represent the conditioning for the music generation. During training, MusicLM uses the MuLan music embedding, while at inference time, the MuLan text embedding is used. \\
The discrete audio representations from SoundStream and the discrete text representations from MuLan are used in a hierarchical sequence-to-sequence modeling task for text-conditioned music generation. The first stage learns the mapping from the MuLan audio tokens to the semantic tokens. These semantic tokens, derived from pre-trained audio data models, facilitate the representation and modeling of extensive musical structures. The second stage predicts the acoustic tokens conditioned on both MuLan audio and semantic tokens. Decoder-only Transformers are used for modeling both stages, and the acoustic modeling stage is further split into coarse and fine modeling stages to avoid long token sequences. \\
During training, SoundStream and w2vBERT are trained on the Free Music Archive (FMA) dataset\footnote{https://github.com/mdeff/fma}. In contrast, the tokenizers and autoregressive models for the semantic and acoustic modeling stages are trained on a large dataset containing five million audio clips. The stages are trained with multiple passes over the training data, using 30 and 10-second random crops of the target audio for the semantic and acoustic stages, respectively. During inference, the MuLan text embedding is used as the conditioning signal, and temperature sampling is used for the autoregressive sampling in all stages. The temperature values are chosen based on subjective inspection to provide a good trade-off between the diversity and temporal consistency of the generated music.
\subsubsection{Features and Capabilities } The MusicLM website showcases the ability of an AI-powered music generation model to produce audio based on concise textual descriptions, including information on music genre, instrument, tempo, and emotion.
The model has demonstrated its potential in generating both short (30 seconds) and long (5 minutes) music pieces. Additionally, the model offers a story mode feature, where it generates music based on a sequence of text prompts that convey changes in mood or plot. This feature demonstrates the model's capability to seamlessly transition
between different levels of mood in its generated music. Furthermore, the model exhibits flexibility in its conditioning, allowing for the generation of music from a provided melody consistent with the provided text description, as well as from paintings with accompanying captions. The model also offers the ability to generate music for different levels of musician experience.
\subsubsection{Advantages and Limitations } MusicLM model has garnered significant attention due to its ability to produce music that can last for minutes, and its high audio quality performance. One of the main advantages of MusicLM is that it can generate complex music that has a structure and coherence similar to that of music created by human composers. This means that the model has the potential to be used for a variety of applications, from generating background music for videos and games to assisting music composers in the creative process. \\
Another advantage of MusicLM is its ability to generate long pieces of music, which is not always possible with other music generation models. This is particularly useful for applications such as creating background music that requires a continuous flow of music for an extended period. Additionally, the model can produce high-quality audio, which is crucial for ensuring that the generated music meets the desired standards. \\
However, the significant limitation of MusicLM is that the official model and checkpoints have not been released as an open-source project due to royalty risks.

\subsection{MusicGen}
\subsubsection{Overview}
MusicGen \cite{MusicGen2023}, a component of the Audiocraft \cite{copet2023simple} library, is a recent music generation model developed by using the transformer architecture. This model generates music samples based on textual descriptions or melodic features, thereby establishing a unified framework for language and music comprehension. Its potential applications encompass various areas, including music composition and soundtrack creation.

\subsubsection{Architecture}
MusicGen is a single-stage auto-regressive Transformer model \cite{MusicGen2023}, trained over a 32kHz EnCodec tokenizer with four codebooks sampled at 50 Hz. It differs from methods such as MusicLM by not requiring a self-supervised semantic representation, and generating all four codebooks simultaneously. By incorporating a small delay between the codebooks, parallel prediction of the codebooks is possible, reducing the auto-regressive steps to 50 per second of audio.\\
It includes a Text conditioning, which is performed by calculating a conditioning tensor from a matching textual description for the input audio. Different text representation methods are utilized, including pretrained text encoders such as T5 \cite{raffel2020exploring}, instruct-based language model FLAN-T5 \cite{chung2022scaling}, and joint text-audio representation CLAP \cite{wu2023largescale}.\\
Its melody conditioning adopts an iterative refinement approach, controlling the melodic structure by conditioning on the input's chromagram and text description. An information bottleneck, introduced by selecting the dominant time-frequency bin at each timestep, helps prevent overfitting and eliminates the need for supervised data.

\subsubsection{Functionality and Capabilities}
MusicGen enables users to generate music samples based on textual descriptions or melodic features. It offers control over the generation process, catering to different musical preferences and requirements. The model supports various music styles and can handle different levels of abstraction in textual descriptions while allowing for melody adjustments during the generation process.\\
MusicGen includes four pre-trained models: small (300M), medium (1.5B), melody (1.5B, with text-to-music and text+melody-to-music capabilities), and large (3.3B, text-to-music only). This range of models enables users to choose the model that best suits their specific use cases and resource constraints.\\
MusicGen was selected and seamlessly integrated as the music generation component of AudioCraft\cite{copet2023simple}, an AI audio toolkit. 
Implemented interfaces to interact with MusicGen are availble through a Jupyter notebook  \footnote{https://github.com/facebookresearch/audiocraft/blob/main/demo.ipynb}, local Gradio \footnote{https://colab.research.google.com/drive/1fxGqfg96RBUvGxZ1XXN07s3DthrKUl4-?usp=sharing}, and HuggingFace Space  \footnote{https://huggingface.co/spaces/facebook/MusicGen}. To operate the model locally, a GPU is required, with a recommendation of 16GB memory, but smaller GPUs can still generate shorter sequences or operate with the small model.

\subsubsection{Advantages and Limitations }
MusicGen stands out as a state-of-the-art single-stage controllable music generation model, with the ability to be conditioned on both text and melody. It uses simple codebook interleaving strategies to generate high-quality music, which also results in a reduction in the number of autoregressive time steps compared to the more traditional flattening approach. Furthermore, MusicGen's performance has been thoroughly analyzed across different model sizes, conditioning methods, and text pre-processing techniques, demonstrating its versatility \cite{MusicGen2023}.

In the domain of text-to-music generation, MusicGen has shown superiority over other models such as Mousai\cite{schneider2023mousai}, Riffusion\cite{Forsgren_Martiros_2022}, MusicLM\cite{MusicLM_paper}, and Noise2Music\cite{huang2023noise2music}. Its high-quality audio output exhibits better adherence to the provided text description. When it comes to melody generation, MusicGen effectively generates music conditioned on a given melody, even when chroma is dropped at inference time, a robustness attribute that further distinguishes it.

MusicGen has received high marks on both objective and subjective evaluation metrics, including the Fréchet Audio Distance (FAD)\cite{kilgour2019frechet}, Kullback-Leibler Divergence (KL), and the CLAP score\cite{wu2023largescale, huang2023makeanaudio}. These quantitative findings are bolstered by qualitative assessments from human listeners who rate the audio quality and relevance to the text input highly.

However, MusicGen does have some limitations. Its generation method doesn't offer fine-grained control over adherence of the generated music to the conditioning; this largely depends on the conditioning framework (CF). Additionally, while data augmentation for text conditioning is straightforward, audio conditioning requires more investigation concerning data augmentation methods, types, and the amount of guidance necessary.

Ethically, MusicGen presents certain challenges similar to many large-scale generative models. The training dataset used has a significant proportion of Western-style music, which may result in a potential lack of diversity in the generated music. However, through the simplification of its design, such as using a single-stage language model and a reduced number of auto-regressive steps, the model hopes to expand its application to new datasets.

\section{Visual-based Music Generation Tools}

Visual-based Music Generation tools signify a novel trend in the field, utilizing visual inputs like images or videos to generate musical pieces. These models transcend traditional music generation boundaries, creating an auditory reflection of visual stimuli. As part of the larger parameter-free models category, they generate music autonomously from the given visual prompts, offering unique possibilities in the music generation landscape.

\subsection{Controllable Music Transformer}

\subsubsection{Overview } Controllable Music Transformer (CMT)\cite{Di_2021} is a transformer-based approach that specializes in generating background music that matches the given video in terms of rhythm and mood. It proposes a novel rhythmic relation between video and music, connecting timing, motion speed, and motion saliency from video with beat, simu-note density, and simu-note strength from music, respectively. In this way, CMT can generate background music that matches the given video in terms of rhythm and mood.

\subsubsection{Architecture } CMT uses a transformer-based approach with 12 self-attention layers and eight attention heads to analyze the rhythm of the video and the music and establish connections between them. The model is trained on the Lakh Pianoroll Dataset \cite{Dong2018MuseGAN, Raffel2016AudioMIDI}, which contains over 174,000 multi-track pianorolls used to represent MIDI music. The CMT model combines rhythmic features extracted from both the video and the MIDI music, which are represented as compound words, and embeds them with beat-timing encoding to create a sequence of music tokens. The transformer model then predicts the attributes of each token, such as pitch, duration, and instrument type, to generate the final music. Finally, in the inference stage, CMT replaces the rhythmic features with those from the video to create music that matches the video's rhythm.

\subsubsection{Features and Capabilities } CMT can generate background music that matches the given video in terms of rhythm and mood. It allows local control of rhythmic features and global control of music genres and instruments. After finishing training, the CMT model has already understood the meaning behind strength and density attributes, and thus we only need to replace those two attributes when appropriate in the inference stage to make the music more harmonious with the given video. CMT introduces a hyperparameter to control the compatibility between music and video, making the generated music more suitable than human-made music. Furthermore, CMT introduces beat-timing encoding to leverage time or length information from the video and genre/instrument type selection to choose different initial tokens in the inference stage.

\textit{Advantages and Limitations }The proposed CMT model has demonstrated satisfactory performance\cite{Di_2021} in generating background music that closely aligns with the rhythm and mood of the given video. It accomplishes this by controlling the density and strength attributes of the music, which contributes to creating a more harmonious music-video combination. Additionally, CMT introduces a hyper-parameter that allows users to control the degree of compatibility between the music and video. However, it is essential to note that generating background music for videos longer than two minutes can be computationally expensive, which may limit its applicability to certain use cases. Furthermore, while CMT produces highly compatible music with the video, the melodiousness of the generated music still falls short of the music in the training set.

\subsection{V-MusProd: Video Music Production AI Model}

\subsubsection{Overview} V-MusProd \cite{zhuo2022video} is a model tailored to generate a fitting musical composition for a given video, leveraging a unique decoupling method for chords, melody, and accompaniment. This approach extracts and translates semantic, color, and motion features from the video to guide the music generation process, ensuring a congruous relationship between the video's visual content and the AI-generated music.
\subsubsection{Architecture } V-MusProd, our novel music generation framework, is composed of a video controller and a music generator. The video controller extracts visual and rhythmic features, which then serve as the contextual input for the music generator. The music generator follows a decoupling process for generating music, with three independently trained stages: Chord, Melody, and Accompaniment. The final music piece is composed by merging the melody and accompaniment tracks generated at the inference time.

The video controller employs meaningful feature extraction techniques from the video to simplify the learning process. We extract semantic, color, and motion features separately to guide the music generation model. Semantic features are extracted using a pretrained CLIP2Video model to encode raw video frames into semantic feature tokens. Color features, representing the color distribution in a non-linear manifold, serve as control signals for chord generation. Motion features, determined by computing RGB differences, are used to ascertain the music tempo.

Semantic and color features are processed through separate transformer encoders and then concatenated at the length dimension. A learnable embedding is added to distinguish between color feature or semantic feature tokens, which are then fed into a transformer encoder for inter-modality and temporal fusion. The fused output serves as keys and values of cross-attention in the Chord Transformer.

The music generator, consisting of a Chord Transformer, a Melody Transformer, and an Accompaniment Transformer, is designed to generate symbolic music conditioned on the extracted video feature.

Chord Transformer adopts a transformer decoder architecture to learn the long-term dependency of input video feature sequences. Melody Transformer, following an encoder-decoder transformer architecture, generates a note sequence as the output melody based on the input chord sequence. Accompaniment Transformer, similarly utilizing an encoder-decoder transformer, generates the accompaniment sequence based on both the chords and melody. The final music piece is formed by merging the generated accompaniment with the melody.

\subsubsection{Features and Capabilities }

V-MusProd represents a significant step forward in video-conditional music generation. It combines the visual cues from video content and music generation in an unprecedented way, giving the created music a rich and contextually grounded feel.

V-MusProd's distinct attribute is the partitioning of musical elements into chords, melody, and accompaniment. This approach allows for in-depth manipulation and control over each individual component, enhancing the flexibility and precision of the music generation process.

V-MusProd also leverages a diverse range of features from the video content itself. It employs semantic, color, and motion attributes to develop a strong association between the video content and the generated music. This ensures that the produced music is not just a standalone output but rather an audible representation of the video input.

Furthermore, the model's architecture is designed to adapt to an unconditional setting, extending its capabilities beyond just video-conditional music generation. This highlights the model's versatility and potential for broader applications.

\subsubsection{Advantages and Limitations}

V-MusProd showcases substantial improvements over previous video music generation models\cite{foley_music, Koepke20, Su2020, suris2022its, zhu2022quantized}. These earlier models either primarily rely on video rhythmic features or are limited to specific video types. However, V-MusProd demonstrates broader applicability. It can generate music directly from video inputs without the need for paired video-music data or additional annotations, marking a significant advancement in the field. This capability not only simplifies the generation process but also stands as a testament to the model's innovative and robust design.

Objective evaluations highlight the enhanced performance of V-MusProd over CMT \cite{Di_2021} in terms of video-music correspondence and music quality. Subjective evaluations, encompassing a user study involving music composition experts and non-experts, also substantiate this finding, with V-MusProd outperforming CMT in nearly all metrics.

However, V-MusProd has some limitations. Currently, the dataset and the music generated by V-MusProd contain only piano tracks, indicating a lack of exploration of other instruments such as drums, guitars, and strings. Furthermore, the model does not explicitly account for high-level emotional features or repetitive structures of music phrases, a potential area for further improvement. The three stages of V-MusProd's method are trained separately instead of end-to-end, a factor that might impact the model's overall performance and efficiency. Despite these limitations, V-MusProd serves as a promising baseline for future research in the field of video-conditional music generation.

\subsection{Foley Music} 
\subsubsection{Overview } Foley Music\cite{foley_music} is a tool that generates music that matches the visual content of a video clip. The tool uses a deep neural network to learn the relationship between the audio effects and the music that it generates. 

\subsubsection{Architecture } In this work, the authors propose generating music from videos by identifying two crucial intermediate representations: body key points and MIDI events. The generation process is formulated as a motion-to-MIDI translation problem, and a Graph-Transformer framework is presented to predict MIDI event sequences that correspond to the body movements accurately. The proposed system comprises three main components: a visual encoder, a MIDI decoder, and an audio synthesizer. To begin with, the visual encoder takes the video frames as input and extracts keypoint coordinates using a Graph Convolutional Network (GCN). This helps to capture the body dynamics and produce a latent representation over time. The MIDI decoder then takes this video sequence representation and generates a sequence of MIDI events. Lastly, the generated MIDI event sequence is converted to the corresponding waveform with an off-the-shelf music synthesizer tool. 

\subsubsection{Features and Capabilities } The Foley Music system is built on PyTorch, a popular deep learning framework, and is capable of generating music clips for a variety of instruments including accordion, bass, bassoon, cello, guitar, piano, tuba, ukulele, and violin. To prepare the audio data for music generation, MIDI events are extracted from audio recordings. The system's performance is demonstrated in an experiment that trains it on a 6-second video clip from a large-scale music performance video dataset. The training process requires the use of optimizers and regularizers to ensure high-quality results. As mentioned, a software synthesizer is required to obtain the final generated music waveforms. 

\subsubsection{Advantages and Limitations } Experimental results reveal that the model is capable of working effectively on various types of music performance videos. The results indicate that the framework successfully establishes correlations between visual and music signals using body keypoints and MIDI representations. \\
Notably, the MIDI representations used in the framework are transparent and fully interpretable, providing flexibility in music editing. MIDI representation can be conveniently modified to generate music in diverse styles by utilizing the MIDI representations. However, this requires additional music synthesizers to render the music, as the framework does not include a neural synthesizer.

\section{Commercial Music Generation Tools}\label{commercial-music-generation-tools}

The commercial market is flooded with a variety of music generation tools that cater to users who desire to create music easily and quickly, but might not obtain prior musical knowledge or coding skills. These tools are typically equipped with web-based user interfaces allowing users to manipulate parameters such as emotion, tempo, and length to generate music according to their preferences. However, these tools can be broadly categorized based on their user interface and interaction model: Pre-set Parameter Tools, Interactive Generation Tools, and Auto-generation Tools.

Parameter-based models include platforms that allow users to select various parameters such as mood, genre, or activity as the guiding elements for music generation. Some platforms in this category offer a more advanced layer of customization, enabling users to adjust the structure and components of the generated music, such as  Mubert\footnote{\url{https://mubert.com}}, Boomy\footnote{\url{https://boomy.com}}, Ecrett Music\footnote{\url{https://ecrettmusic.com}}, and Soundraw.io\footnote{\url{https://soundraw.io}}.

Prompt-based models, such as SongR\footnote{\url{https://app.songr.ai}}, offer a unique interaction where users can input their lyrics or melodic ideas to guide the music generation process.

Style-driven models adapt the composition process based on user-provided influences, which can include an existing music file or a specific emotional impact. Additionally, these models may also offer preset styles to guide the composition, such as Aiva.ai\footnote{\url{https://www.aiva.ai}}.

Despite their ease of use and accessibility, these tools often do not provide detailed information about their underlying model architecture, leaving users in the dark about the methodologies or algorithms behind the music they generate. This lack of transparency, coupled with potentially limited customization options, could mean these tools fall short for users with more specialized or complex musical needs.

Nonetheless, commercial music generation tools hold value for their ability to quickly and conveniently generate music. The absence of in-depth model descriptions, however, renders a comprehensive evaluation of these tools' performance or effectiveness challenging. While this paper will not delve into the specifics of commercial music models, their contribution to making music generation more accessible and convenient for users is recognized.

\section{Conclusion}
In this survey we described models that allow music generated on parameters, prompts, and video clips. Our analysis highlights each tool's unique advantages and limitations, such as flexibility, sophistication, and quality of the generated music. For example, some tools provide flexibility in developing MIDI files but require additional software synthesizers to process the music. Other tools (especially the prompt-based diffusion models) generate sophisticated music but need more flexibility in generating music with different instruments. A challenge remains the ability to generate longer pieces with a good musical pattern. We also acknowledge that this review is not exhaustive due to the rapidly evolving nature of this field. Nonetheless, the range of AI music generation tools available shows promise in revolutionizing the music industry, enhancing creativity, and broadening the range of musical expression. We anticipate the emergence of more sophisticated models will overcome the current limitations and provide more flexible, user-friendly, and high-quality AI music generation tools.

\bibliographystyle{plain}
\bibliography{reference}

\begin{thebibliography}{100}

\bibitem{abcnotation}
{ABC Notation}.
\newblock {ABC Notation}.
\newblock \url{https://abcnotation.com/}, Accessed 2023.

\bibitem{bitmidi}
Feross Aboukhadijeh.
\newblock Bitmidi: Free midi files, 2023.
\newblock Accessed: 2023-05-15.

\bibitem{MusicLM_paper}
Andrea Agostinelli, Timo~I. Denk, Zalán Borsos, Jesse Engel, Mauro Verzetti,
  Antoine Caillon, Qingqing Huang, Aren Jansen, Adam Roberts, Marco
  Tagliasacchi, Matt Sharifi, Neil Zeghidour, and Christian Frank.
\newblock Musiclm: Generating music from text, 2023.

\bibitem{google_ai_start}
Google AI.
\newblock Getting started, n.d.

\bibitem{google_ai}
Google AI.
\newblock Magenta.
\newblock \url{https://magenta.tensorflow.org/}, n.d.
\newblock Accessed 10 Feburary 2023.

\bibitem{MarkovProcess}
Charles Ames.
\newblock The markov process as a compositional model: A survey and tutorial.
\newblock {\em Leonardo}, 22(2):175--187, 1989.

\bibitem{Biles1994}
John~A Biles.
\newblock Genjam: A genetic algorithm for generating jazz solos.
\newblock {\em Proceedings of the international computer music conference},
  pages 131--137, 1994.

\bibitem{bouchard_musiclm-pytorch_2023}
Louis Bouchard.
\newblock Musiclm - pytorch.
\newblock \url{https://github.com/lucidrains/musiclm-pytorch}, February 2023.
\newblock Accessed: March 15, 2023.

\bibitem{Briot2020}
Jean-Pierre Briot and Fran\c{c}ois Pachet.
\newblock Deep learning for music generation: Challenges and directions.
\newblock {\em Neural Comput. Appl.}, 32(4):981–993, feb 2020.

\bibitem{Carnovalini_2020}
Filippo Carnovalini and Antonio Rodà.
\newblock Computational creativity and music generation systems: An
  introduction to the state of the art.
\newblock {\em Frontiers in Artificial Intelligence}, 3, 2020.

\bibitem{sparse-transformer}
Rewon Child, Scott Gray, Alec Radford, and Ilya Sutskever.
\newblock Generating long sequences with sparse transformers, 2019.

\bibitem{chung2022scaling}
Hyung~Won Chung, Le~Hou, Shayne Longpre, Barret Zoph, Yi~Tay, William Fedus,
  Yunxuan Li, Xuezhi Wang, Mostafa Dehghani, Siddhartha Brahma, Albert Webson,
  Shixiang~Shane Gu, Zhuyun Dai, Mirac Suzgun, Xinyun Chen, Aakanksha
  Chowdhery, Alex Castro-Ros, Marie Pellat, Kevin Robinson, Dasha Valter,
  Sharan Narang, Gaurav Mishra, Adams Yu, Vincent Zhao, Yanping Huang, Andrew
  Dai, Hongkun Yu, Slav Petrov, Ed~H. Chi, Jeff Dean, Jacob Devlin, Adam
  Roberts, Denny Zhou, Quoc~V. Le, and Jason Wei.
\newblock Scaling instruction-finetuned language models, 2022.

\bibitem{chung2021w2vbert}
Yu-An Chung, Yu~Zhang, Wei Han, Chung-Cheng Chiu, James Qin, Ruoming Pang, and
  Yonghui Wu.
\newblock W2v-bert: Combining contrastive learning and masked language modeling
  for self-supervised speech pre-training, 2021.

\bibitem{MusicGen2023}
Jade Copet, Felix Kreuk, Itai Gat, Tal Remez, David Kant, Gabriel Synnaeve,
  Yossi Adi, and Alexandre Défossez.
\newblock Simple and controllable music generation, 2023.

\bibitem{copet2023simple}
Jade Copet, Felix Kreuk, Itai Gat, Tal Remez, David Kant, Gabriel Synnaeve,
  Yossi Adi, and Alexandre Défossez.
\newblock Simple and controllable music generation.
\newblock {\em arXiv preprint arXiv:2306.05284}, 2023.

\bibitem{covach2005what}
John Covach and Andrew Flory.
\newblock {\em What's that sound? An introduction to rock and its history}.
\newblock WW Norton \& Company, 2005.

\bibitem{shu28011}
Luisa~Rocha de~Azevedo~Santos, Carlos~Silla Jr., and Marjory Da~Costa Abreu.
\newblock A methodology for procedural piano music composition with mood
  templates using genetic algorithms.
\newblock In {\em 11th International Conference of Pattern Recognition Systems
  (ICPRS 2021)}, pages 1--6. IET, October 2021.

\bibitem{denisova2017evolution}
Elena Denisova-Schmidt.
\newblock {\em The Evolution of Music: Culture, Technology, and Society}.
\newblock Routledge, 2017.

\bibitem{magentaImprovRNN}
Magenta Developers.
\newblock Magenta improv rnn.
\newblock
  \url{https://github.com/magenta/magenta/blob/main/magenta/models/improv_rnn/README.md},
  2021.
\newblock Accessed on April 17, 2023.

\bibitem{magentaMelodyRNN}
Magenta Developers.
\newblock Magenta melody rnn.
\newblock
  \url{https://github.com/magenta/magenta/blob/main/magenta/models/melody_rnn/README.md},
  2021.
\newblock Accessed on April 17, 2023.

\bibitem{magentaPolyphonyRNN}
Magenta Developers.
\newblock Magenta polyphony rnn.
\newblock
  \url{https://github.com/magenta/magenta/blob/main/magenta/models/polyphony_rnn/README.md},
  2021.
\newblock Accessed on April 17, 2023.

\bibitem{magenta-models}
Magenta Developers.
\newblock Magenta model directory.
\newblock \url{https://github.com/magenta/magenta/tree/main/magenta/models},
  2023.

\bibitem{jukebox}
Prafulla Dhariwal, Heewoo Jun, Christine Payne, Jong~Wook Kim, Alec Radford,
  and Ilya Sutskever.
\newblock Jukebox: A generative model for music, 2020.

\bibitem{Di_2021}
Shangzhe Di, Zeren Jiang, Si~Liu, Zhaokai Wang, Leyan Zhu, Zexin He, Hongming
  Liu, and Shuicheng Yan.
\newblock Video background music generation with controllable music
  transformer.
\newblock In {\em Proceedings of the 29th {ACM} International Conference on
  Multimedia}. {ACM}, oct 2021.

\bibitem{Dong2018MuseGAN}
Hao-Wen Dong, Wen-Yi Hsiao, Li-Chia Yang, and Yi-Hsuan Yang.
\newblock Musegan: Multi-track sequential generative adversarial networks for
  symbolic music generation and accompaniment.
\newblock In {\em Proceedings of the 32nd AAAI Conference on Artificial
  Intelligence (AAAI)}, 2018.

\bibitem{dubreuil2020hands}
A.~DuBreuil.
\newblock {\em Hands-On Music Generation with Magenta: Explore the role of deep
  learning in music generation and assisted music composition}.
\newblock Packt Publishing, 2020.

\bibitem{engel2018gansynth}
Jesse Engel, Kumar~Krishna Agrawal, Shuo Chen, Ishaan Gulrajani, Chris Donahue,
  and Adam Roberts.
\newblock {GANS}ynth: Adversarial neural audio synthesis.
\newblock In {\em International Conference on Learning Representations}, 2019.

\bibitem{engel2017neural}
Jesse Engel, Cinjon Resnick, Adam Roberts, Sander Dieleman, Douglas Eck, Karen
  Simonyan, and Mohammad Norouzi.
\newblock Neural audio synthesis of musical notes with wavenet autoencoders,
  2017.

\bibitem{Farzaneh2019}
Majid Farzaneh and Rahil~Mahdian Toroghi.
\newblock Music generation using an interactive evolutionary algorithm.
\newblock In {\em Pattern Recognition and Artificial Intelligence}, pages
  207--217. Springer International Publishing, December 2019.

\bibitem{ferreira2021learning}
Lucas~N. Ferreira and Jim Whitehead.
\newblock Learning to generate music with sentiment, 2021.

\bibitem{Forsgren_Martiros_2022}
Seth* Forsgren and Hayk* Martiros.
\newblock {Riffusion - Stable diffusion for real-time music generation}, 2022.

\bibitem{foley_music}
Chuang Gan, Deng Huang, Peihao Chen, Joshua~B. Tenenbaum, and Antonio Torralba.
\newblock Foley music: Learning to generate music from videos.
\newblock In {\em Computer Vision – ECCV 2020: 16th European Conference,
  Glasgow, UK, August 23–28, 2020, Proceedings, Part XI}, page 758–775,
  Berlin, Heidelberg, 2020. Springer-Verlag.

\bibitem{Gautam2018MusicCW}
Sudhanshu Gautam and Sarita Soni.
\newblock Music composition with artificial intelligence system based on markov
  chain and genetic algorithm.
\newblock 2018.

\bibitem{groove2019}
Jon Gillick, Adam Roberts, Jesse Engel, Douglas Eck, and David Bamman.
\newblock Learning to groove with inverse sequence transformations.
\newblock In {\em International Conference on Machine Learning (ICML)}, 2019.

\bibitem{NIPS2014_5ca3e9b1}
Ian Goodfellow, Jean Pouget-Abadie, Mehdi Mirza, Bing Xu, David Warde-Farley,
  Sherjil Ozair, Aaron Courville, and Yoshua Bengio.
\newblock Generative adversarial nets.
\newblock In Z.~Ghahramani, M.~Welling, C.~Cortes, N.~Lawrence, and K.Q.
  Weinberger, editors, {\em Advances in Neural Information Processing Systems},
  volume~27. Curran Associates, Inc., 2014.

\bibitem{GUDIVADA2015203}
Venkat~N. Gudivada, Dhana Rao, and Vijay~V. Raghavan.
\newblock Chapter 9 - big data driven natural language processing research and
  applications.
\newblock In Venu Govindaraju, Vijay~V. Raghavan, and C.R. Rao, editors, {\em
  Big Data Analytics}, volume~33 of {\em Handbook of Statistics}, pages
  203--238. Elsevier, 2015.

\bibitem{maestro-dataset}
Curtis Hawthorne, Andriy Stasyuk, Adam Roberts, Ian Simon, Cheng-Zhi~Anna
  Huang, Sander Dieleman, Erich Elsen, Jesse Engel, and Douglas Eck.
\newblock The maestro dataset.
\newblock arXiv preprint arXiv:1810.12247, 2018.

\bibitem{hernandezolivan2021music}
Carlos Hernandez-Olivan and Jose~R. Beltran.
\newblock Music composition with deep learning: A review, 2021.

\bibitem{SimoneHill2023}
Simone Hill.
\newblock Markov melody generator.
\newblock {\em University of Massachusetts Lowell}, 2023.

\bibitem{hsiao2021compound}
Wen-Yi Hsiao, Jen-Yu Liu, Yin-Cheng Yeh, and Yi-Hsuan Yang.
\newblock Compound word transformer: Learning to compose full-song music over
  dynamic directed hypergraphs, 2021.

\bibitem{huang2018music}
Cheng-Zhi~Anna Huang, Ashish Vaswani, Jakob Uszkoreit, Noam Shazeer, Ian Simon,
  Curtis Hawthorne, Andrew~M. Dai, Matthew~D. Hoffman, Monica Dinculescu, and
  Douglas Eck.
\newblock Music transformer, 2018.

\bibitem{Mulan2022}
Qingqing Huang, Aren Jansen, Joonseok Lee, Ravi Ganti, Judith~Yue Li, and
  Daniel P.~W. Ellis.
\newblock Mulan: A joint embedding of music audio and natural language.
\newblock In {\em Proceedings of the the 23rd International Society for Music
  Information Retrieval Conference (ISMIR)}, 2022.

\bibitem{huang2022mulan}
Qingqing Huang, Aren Jansen, Joonseok Lee, Ravi Ganti, Judith~Yue Li, and
  Daniel~PW Ellis.
\newblock Mulan: A joint embedding of music audio and natural language.
\newblock {\em arXiv preprint arXiv:2208.12415}, 2022.

\bibitem{huang2023noise2music}
Qingqing Huang, Daniel~S. Park, Tao Wang, Timo~I. Denk, Andy Ly, Nanxin Chen,
  Zhengdong Zhang, Zhishuai Zhang, Jiahui Yu, Christian Frank, Jesse Engel,
  Quoc~V. Le, William Chan, Zhifeng Chen, and Wei Han.
\newblock Noise2music: Text-conditioned music generation with diffusion models,
  2023.

\bibitem{huang2023makeanaudio}
Rongjie Huang, Jiawei Huang, Dongchao Yang, Yi~Ren, Luping Liu, Mingze Li,
  Zhenhui Ye, Jinglin Liu, Xiang Yin, and Zhou Zhao.
\newblock Make-an-audio: Text-to-audio generation with prompt-enhanced
  diffusion models, 2023.

\bibitem{9054554}
Junyan Jiang, Gus~G. Xia, Dave~B. Carlton, Chris~N. Anderson, and Ryan~H.
  Miyakawa.
\newblock Transformer vae: A hierarchical model for structure-aware and
  interpretable music representation learning.
\newblock In {\em ICASSP 2020 - 2020 IEEE International Conference on
  Acoustics, Speech and Signal Processing (ICASSP)}, pages 516--520, 2020.

\bibitem{kilgour2019frechet}
Kevin Kilgour, Mauricio Zuluaga, Dominik Roblek, and Matthew Sharifi.
\newblock Fr\'echet audio distance: A metric for evaluating music enhancement
  algorithms, 2019.

\bibitem{kingma2022autoencoding}
Diederik~P Kingma and Max Welling.
\newblock Auto-encoding variational bayes, 2022.

\bibitem{Koepke20}
A.S. Koepke, O.~Wiles, Y.~Moses, and A.~Zisserman.
\newblock Sight to sound: An end-to-end approach for visual piano
  transcription.
\newblock In {\em International Conference on Acoustics, Speech, and Signal
  Processing}, 2020.

\bibitem{krause2017}
Ben Krause, Iain Murray, Steve Renals, and Liang Lu.
\newblock Multiplicative {LSTM} for sequence modelling.
\newblock {\em ICLR Workshop track}, 2017.

\bibitem{laitz2012complete}
Steven~G Laitz.
\newblock {\em The complete musician: An integrated approach to tonal theory,
  analysis, and listening}.
\newblock Oxford University Press, 2012.

\bibitem{lecun2015deep}
Yann LeCun, Yoshua Bengio, and Geoffrey Hinton.
\newblock Deep learning.
\newblock {\em Nature}, 521(7553):436--444, 2015.

\bibitem{leider2004digital}
Colby Leider.
\newblock {\em Digital audio workstation}.
\newblock McGraw-Hill New York, 2004.

\bibitem{classicalarchives}
Classical~Archives LLC.
\newblock Classical archives: The largest classical music site in the world,
  2023.
\newblock Accessed: 2023-05-15.

\bibitem{london2004hearing}
Justin London.
\newblock {\em Hearing in time: Psychological aspects of musical meter}.
\newblock Oxford University Press, 2004.

\bibitem{Louie_2020}
Ryan Louie, Andy Coenen, Cheng~Zhi Huang, Michael Terry, and Carrie~J. Cai.
\newblock Novice-ai music co-creation via ai-steering tools for deep generative
  models.
\newblock In {\em Proceedings of the 2020 CHI Conference on Human Factors in
  Computing Systems}, CHI '20, page 1–13, New York, NY, USA, 2020.
  Association for Computing Machinery.

\bibitem{magenta_github}
Magenta.
\newblock Magenta/magenta: Magenta: Music and art generation with machine
  intelligence, n.d.

\bibitem{marques2000music}
Manuel Marques, V~Oliveira, S~Vieira, and AC~Rosa.
\newblock Music composition using genetic evolutionary algorithms.
\newblock In {\em Proceedings of the 2000 Congress on Evolutionary Computation.
  CEC00 (Cat. No. 00TH8512)}, volume~1, pages 714--719. IEEE, 2000.

\bibitem{meyer1989style}
Leonard~B Meyer.
\newblock {\em Style and music: Theory, history, and ideology}.
\newblock University of Pennsylvania Press, 1989.

\bibitem{moser2009engineering}
Michael M{\"o}ser.
\newblock Engineering acoustics.
\newblock {\em Nova York (Estados Unidos): Springer Publishing}, 2009.

\bibitem{LSTM_intro}
Christopher Olah.
\newblock Understanding lstm networks, 2015.

\bibitem{openai_2020}
OpenAI, Apr 2020.

\bibitem{palatucci2009zero}
Mark Palatucci, Dean Pomerleau, Geoffrey~E Hinton, and Tom~M Mitchell.
\newblock Zero-shot learning with semantic output codes.
\newblock {\em Advances in neural information processing systems}, 22, 2009.

\bibitem{openai-musenet}
Christine Payne.
\newblock Musenet.
\newblock \url{https://openai.com/blog/musenet/}, 2019.
\newblock Accessed: 2023-05-15.

\bibitem{piston1987harmony}
Walter Piston.
\newblock {\em Harmony}.
\newblock WW Norton \& Company, 1987.

\bibitem{DMAE}
Konpat Preechakul, Nattanat Chatthee, Suttisak Wizadwongsa, and Supasorn
  Suwajanakorn.
\newblock Diffusion autoencoders: Toward a meaningful and decodable
  representation.
\newblock In {\em 2022 IEEE/CVF Conference on Computer Vision and Pattern
  Recognition (CVPR)}, pages 10609--10619, 2022.

\bibitem{radford2017learning}
Alec Radford, Rafal Jozefowicz, and Ilya Sutskever.
\newblock Learning to generate reviews and discovering sentiment.
\newblock {\em arXiv preprint arXiv:1704.01444}, 2017.

\bibitem{Raffel2016-et}
Colin Raffel.
\newblock Learning-based methods for comparing sequences, with applications to
  {audio-to-MIDI} alignment and matching, June 2016.

\bibitem{Raffel2016AudioMIDI}
Colin Raffel.
\newblock {\em Learning-Based Methods for Comparing Sequences, with
  Applications to Audio-to-MIDI Alignment and Matching}.
\newblock PhD thesis, 2016.

\bibitem{raffel2020exploring}
Colin Raffel, Noam Shazeer, Adam Roberts, Katherine Lee, Sharan Narang, Michael
  Matena, Yanqi Zhou, Wei Li, and Peter~J. Liu.
\newblock Exploring the limits of transfer learning with a unified text-to-text
  transformer, 2020.

\bibitem{Ramanto2017MarkovCB}
Adhika~Sigit Ramanto and Nur~Ulfa Maulidevi.
\newblock Markov chain based procedural music generator with user chosen mood
  compatibility.
\newblock {\em International Journal of Asia Digital Art and Design},
  21:19--24, 2017.

\bibitem{ratcliff2004diffusion}
Roger Ratcliff, Pablo G{\'o}mez, and Gail McKoon.
\newblock A diffusion model account of response time and accuracy in a
  brightness discrimination task: Fitting real data and failing to fit fake but
  plausible data.
\newblock {\em Psychological Review}, 111(4):931--959, 2004.

\bibitem{razavi2019generating}
Ali Razavi, Aaron van~den Oord, and Oriol Vinyals.
\newblock Generating diverse high-fidelity images with vq-vae-2, 2019.

\bibitem{musicxml}
{Recordare LLC}.
\newblock {MusicXML}.
\newblock \url{https://www.musicxml.com/}, Accessed 2023.

\bibitem{RISSET1999113}
Jean-Claude Risset and David~L. Wessel.
\newblock 5 - exploration of timbre by analysis and synthesis.
\newblock In Diana Deutsch, editor, {\em The Psychology of Music (Second
  Edition)}, Cognition and Perception, pages 113--169. Academic Press, San
  Diego, second edition edition, 1999.

\bibitem{roberts2019hierarchical}
Adam Roberts, Jesse Engel, Colin Raffel, Curtis Hawthorne, and Douglas Eck.
\newblock A hierarchical latent vector model for learning long-term structure
  in music, 2019.

\bibitem{roederer2013physics}
Juan~G Roederer.
\newblock {\em The physics and psychophysics of music: An introduction}.
\newblock Springer Science \& Business Media, 2013.

\bibitem{stable_diffusion}
Robin Rombach, Andreas Blattmann, Dominik Lorenz, Patrick Esser, and Bj\"orn
  Ommer.
\newblock High-resolution image synthesis with latent diffusion models.
\newblock In {\em Proceedings of the IEEE/CVF Conference on Computer Vision and
  Pattern Recognition (CVPR)}, pages 10684--10695, June 2022.

\bibitem{rombach2022highresolution}
Robin Rombach, Andreas Blattmann, Dominik Lorenz, Patrick Esser, and Björn
  Ommer.
\newblock High-resolution image synthesis with latent diffusion models, 2022.

\bibitem{rossing2002science}
Thomas~D. Rossing, F.~Richard Moore, and Paul~A. Wheeler.
\newblock {\em The Science of Sound}.
\newblock Addison-Wesley, 2002.

\bibitem{russell1980}
James Russell.
\newblock A circumplex model of affect.
\newblock {\em Journal of Personality and Social Psychology}, 39:1161--1178, 12
  1980.

\bibitem{schneider2023mousai}
Flavio Schneider, Zhijing Jin, and Bernhard Schölkopf.
\newblock Mo\^usai: Text-to-music generation with long-context latent
  diffusion, 2023.

\bibitem{MuseScore2011}
Werner Schweer and Others.
\newblock Musescore.
\newblock {\em MuseScore BVBA}, 2011.

\bibitem{sethares2005tuning}
W.A. Sethares.
\newblock {\em Tuning, Timbre, Spectrum, Scale}.
\newblock Springer London, 2005.

\bibitem{Shapiro2021}
Ilana Shapiro and Mark Huber.
\newblock Markov chains for computer music generation.
\newblock {\em Journal of Humanistic Mathematics}, 11(2):167--195, July 2021.

\bibitem{Sherstinsky_2020}
Alex Sherstinsky.
\newblock Fundamentals of recurrent neural network ({RNN}) and long short-term
  memory ({LSTM}) network.
\newblock {\em Physica D: Nonlinear Phenomena}, 404:132306, mar 2020.

\bibitem{RNN_LSTM}
Alex Sherstinsky.
\newblock Fundamentals of recurrent neural network (rnn) and long short-term
  memory (lstm) network.
\newblock {\em Physica D: Nonlinear Phenomena}, 404:132306, 2020.

\bibitem{Melisma2003}
Daniel Sleator and David Temperley.
\newblock Melisma stochastic melody generator.
\newblock {\em Melisma Music Analyzer}, 2003.

\bibitem{sloboda2005exploring}
John~A Sloboda.
\newblock {\em Exploring the musical mind: Cognition, emotion, ability,
  function}.
\newblock Oxford University Press, 2005.

\bibitem{APOPCALEAPS}
Jon Sneyers and Danny De~Schreye.
\newblock Apopcaleaps: Automatic music generation with chrism.
\newblock 10 2010.

\bibitem{pmlr-v37-sohl-dickstein15}
Jascha Sohl-Dickstein, Eric Weiss, Niru Maheswaranathan, and Surya Ganguli.
\newblock Deep unsupervised learning using nonequilibrium thermodynamics.
\newblock In Francis Bach and David Blei, editors, {\em Proceedings of the 32nd
  International Conference on Machine Learning}, volume~37 of {\em Proceedings
  of Machine Learning Research}, pages 2256--2265, Lille, France, 07--09 Jul
  2015. PMLR.

\bibitem{gpt-2}
Irene Solaiman, Miles Brundage, Jack Clark, Amanda Askell, Ariel Herbert-Voss,
  Jeff Wu, Alec Radford, Gretchen Krueger, Jong~Wook Kim, Sarah Kreps, Miles
  McCain, Alex Newhouse, Jason Blazakis, Kris McGuffie, and Jasmine Wang.
\newblock Release strategies and the social impacts of language models, 2019.

\bibitem{song2021scorebased}
Yang Song, Jascha Sohl-Dickstein, Diederik~P. Kingma, Abhishek Kumar, Stefano
  Ermon, and Ben Poole.
\newblock Score-based generative modeling through stochastic differential
  equations, 2021.

\bibitem{rule-based-generation}
Randall~Richard Spangler.
\newblock {\em Rule-based analysis and generation of music}.
\newblock PhD thesis, 1999.

\bibitem{Su2020}
Kun Su, Xiulong Liu, and Eli Shlizerman.
\newblock Audeo: Audio generation for a silent performance video.
\newblock In H.~Larochelle, M.~Ranzato, R.~Hadsell, M.F. Balcan, and H.~Lin,
  editors, {\em Advances in Neural Information Processing Systems}, volume~33,
  pages 3325--3337. Curran Associates, Inc., 2020.

\bibitem{suris2022its}
Didac Suris, Carl Vondrick, Bryan Russell, and Justin Salamon.
\newblock It's time for artistic correspondence in music and video, 2022.

\bibitem{oord2016wavenet}
Aaron van~den Oord, Sander Dieleman, Heiga Zen, Karen Simonyan, Oriol Vinyals,
  Alex Graves, Nal Kalchbrenner, Andrew Senior, and Koray Kavukcuoglu.
\newblock Wavenet: A generative model for raw audio, 2016.

\bibitem{oord2018neural}
Aaron van~den Oord, Oriol Vinyals, and Koray Kavukcuoglu.
\newblock Neural discrete representation learning, 2018.

\bibitem{van2008visualizing}
Laurens Van~der Maaten and Geoffrey Hinton.
\newblock Visualizing data using t-sne.
\newblock {\em Journal of machine learning research}, 9(11), 2008.

\bibitem{vaswani2017attention}
Ashish Vaswani, Noam Shazeer, Niki Parmar, Jakob Uszkoreit, Llion Jones,
  Aidan~N. Gomez, Lukasz Kaiser, and Illia Polosukhin.
\newblock Attention is all you need, 2017.

\bibitem{wu2023largescale}
Yusong Wu, Ke~Chen, Tianyu Zhang, Yuchen Hui, Taylor Berg-Kirkpatrick, and
  Shlomo Dubnov.
\newblock Large-scale contrastive language-audio pretraining with feature
  fusion and keyword-to-caption augmentation, 2023.

\bibitem{evolutionary_2011}
Ali~Çağatay Yiiksel, Mehmet~Melih Karci, and A.~Şima Uyar.
\newblock Automatic music generation using evolutionary algorithms and neural
  networks.
\newblock In {\em 2011 International Symposium on Innovations in Intelligent
  Systems and Applications}, pages 354--358, 2011.

\bibitem{zeghidour2021soundstream}
Neil Zeghidour, Alejandro Luebs, Ahmed Omran, Jan Skoglund, and Marco
  Tagliasacchi.
\newblock Soundstream: An end-to-end neural audio codec, 2021.

\bibitem{sdmuse}
Chen Zhang, Yi~Ren, Kejun Zhang, and Shuicheng Yan.
\newblock Sdmuse: Stochastic differential music editing and generation via
  hybrid representation, 2022.

\bibitem{zhang_ren_zhang_yan_2022}
Chen Zhang, Yi~Ren, Kejun Zhang, and Shuicheng Yan.
\newblock Stochastic differential music editing and generation via hybrid
  representation, Sep 2022.

\bibitem{mixup}
Hongyi Zhang, Moustapha Cisse, Yann~N Dauphin, and David Lopez-Paz.
\newblock mixup: Beyond empirical risk minimization.
\newblock {\em arXiv preprint arXiv:1710.09412}, 2017.

\bibitem{zhao2022review}
Ziyi Zhao, Hanwei Liu, Song Li, Junwei Pang, Maoqing Zhang, Yi~Qin, Lei Wang,
  and Qidi Wu.
\newblock A review of intelligent music generation systems, 2022.

\bibitem{zhu2022quantized}
Ye~Zhu, Kyle Olszewski, Yu~Wu, Panos Achlioptas, Menglei Chai, Yan Yan, and
  Sergey Tulyakov.
\newblock Quantized gan for complex music generation from dance videos.
\newblock In {\em Proceedings of the European Conference on Computer Vision
  (ECCV)}, 2022.

\bibitem{zhuo2022video}
Le~Zhuo, Zhaokai Wang, Baisen Wang, Yue Liao, Stanley Peng, Chenxi Bao, Miao
  Lu, Xiaobo Li, and Si~Liu.
\newblock Video background music generation: Dataset, method and evaluation,
  2022.

\end{thebibliography}

\end{document}